\documentclass[journal=jacsat,manuscript=article]{achemso}
\usepackage{setspace}
\usepackage{amsmath}
\usepackage{graphicx}
\usepackage{xcolor}
\usepackage{textcomp}
\usepackage{setspace}
\usepackage{amsmath}
\usepackage{graphicx}
\usepackage{xcolor}
\usepackage{textcomp}
\usepackage{tabularx}
\usepackage{array}
\usepackage{makecell}
\usepackage{booktabs}

\newcommand{\cutoffvalue}{0.2}
\newcommand{\zTxp}{zT^{\mathrm{xp}}}
\newcommand{\zTpr}{zT^{\mathrm{pr}}}
\newcommand{\zTxpi}{zT_{i}^{\mathrm{xp}}}
\newcommand{\zTpri}{zT_{i}^{\mathrm{pr}}}

\newcommand{\titlecontentJ}{

A Generalizable Machine Learning Framework for Half-Heusler Thermoelectrics
}

\title{\titlecontentJ}
\author{Shoeb Athar}
\author{Adrien Mecibah}
\author{Philippe Jund}
\email{philippe.jund@umontpellier.fr}
\affiliation{ICGM, Univ Montpellier, CNRS, ENSCM, 34293 Montpellier, France}
\begin{document}
\maketitle

\begin{abstract}

Machine learning (ML) has emerged as a powerful approach for accelerating the 
design and discovery of thermoelectric (TE) materials, but many models suffer 
from limited generalizability due to biased dataset partitioning and 
suboptimal feature selection. Here we present a robust ML workflow for 
predicting the thermoelectric figure of merit ($zT$) in half-Heusler (hH) 
compounds, emphasizing generalizability to previously unseen compositions. We 
introduce a principal component analysis (PCA)-based train/test splitting 
strategy that generates unbiased, representative partitions of the chemical 
space while preventing data leakage in hierarchical materials datasets, and 
integrate Bayesian hyperparameter optimization with a correlation-based, 
$k$-best feature selection strategy across Random Forest, XGBoost, and Neural 
Network models, complemented by the interpretable SISSO symbolic-regression 
method. Relative to the full set of 118 input features, this procedure reduces the test RMSE by
16--27\% under the same train/test partition. SHAP and SISSO analyses consistently identify temperature, the A-site 
dopant concentration, and the A-site heat of vaporization as the dominant 
descriptors of $zT$. We assess the framework's extrapolative capacity
through evaluation on previously unexplored compositions, recovery of 
literature-reported optimal $zT$ values for established hH systems, and 
additional first-principles calculations for two previously unreported 
compositions, CeCoBi and CeIrBi, which exhibit favorable electronic transport 
characteristics. We demonstrate the framework's practical utility through 
high-throughput screening of the doped hH chemical space, comprising 
approximately $2.46 \times 10^6$ elemental substitution combinations across 
631 thermodynamically screened parent matrices, subject to thermodynamic and 
electronic stability constraints, yielding a concise shortlist of 
high-confidence candidates for future investigation. Rather than focusing 
solely on conventional performance metrics such as RMSE or $R^2$, this work 
establishes a generalizable ML framework emphasizing unbiased evaluation, 
robust feature selection, and physically constrained screening for reliable 
data-driven thermoelectric discovery.

    
\end{abstract}

\section{1. Introduction}

The development of efficient thermoelectric (TE) materials which can produce electricity directly from heat through the Seebeck effect is an urgent need in the context of the present environmental crisis. The performance of a TE material is gauged by the dimensionless figure of merit $zT$ given by \cite{goldsmid_2016} $zT = {S^2\sigma T}/{\kappa}$ (where $S$ is the Seebeck coefficient ($\mathrm{V.K^{-1}}$), $T$ is the working temperature ($\mathrm{K}$), $\sigma$ is the electrical conductivity ($\mathrm{S.m^{-1}}$), and $\kappa$ is the thermal conductivity ($\mathrm{W.m^{-1}.K^{-1}}$). A good TE material, therefore, must have high electrical conductivity and Seebeck coefficient – lumped together as $S^2\sigma$ or the TE power factor  – as well as low thermal conductivity. Identifying such materials is a highly challenging task because of the intricate interplay between these properties. Significant research into efficient TE materials has yet to yield widespread commercial success, hindered by the toxicity and scarcity of constituent elements (e.g., Pb, Te, and Ge) and the susceptibility of TE modules to mechanical failure under fluctuating thermal and mechanical stresses \cite{athar_2023,athar_2025b} . Steady progress in artificial intelligence has opened new horizons in the accelerated discovery and design of functional materials \cite{na_chang_2022} versus time-consuming and / or trial-and-error-based experimental or density functional theory (DFT) pathways. Machine learning (ML) has emerged as a powerful technique to design and screen materials with required functionalities \cite{na_chang_2022}. TE materials have also recently benefited from the composition design assisted by ML \cite{katsura_2019}. Several studies have demonstrated that ML can not only be used to predict individual transport properties but also complex properties like the power factor or the $zT$ for various materials \cite{Athar_2026}. However, despite several such studies reporting high predictive performance metrics ($R^2$, RMSE, MAE), few have translated into the discovery and experimental synthesis of new high-performance TE materials \cite{wang_2023, Athar_2026}. This gap persists even with advances in ML techniques, largely because an exhaustive search over vast compositional spaces remains challenging \cite{wang_2023} due to one or more of the following reasons: lack of large curated datasets; poor generalizability of ML models; the structural diversity and complexity of TE materials \cite{na_2021}; and structural/phase stability of the predicted materials. We have addressed the issue of dataset size and curation in our previous work \cite{Athar_m_j_2025}. To mitigate the problems associated with structural diversity and complexity of TE materials, we restrict the present study on a specific structural prototype: half-Heusler (hH) TE materials. While the ideal solution is the inclusion of relevant structural and crystallographic features, the data-diversity in the existing datasets \cite{Athar_m_j_2025} with respect to different structural prototypes is not sufficient to allow for such an approach. The high-throughput synthesis approach of phase- and structurally-stable ML predicted materials will be the subject of our next work. In this work, we focus on the causes of poor generalizability of ML-models for TE materials and the solutions thereof. 

A typical machine learning workflow entails the following modules: dataset construction and curation, data pre-processing, feature selection, model construction and evaluation, model interpretation, and high-throughput screening (HTS). Generalizability of an ML-model particularly depends on the first four components. While several studies have reported sophisticated workflows for ML-application for TE materials \cite{parse_n_2024, li_2022, li_2023, jia_2024, barua_2024} , some key challenges associated with dataset pre-processing (splitting train/test data) and feature selection process have not been addressed yet. The first challenge concerns the splitting of training and test data. Statistical bounds like Hoeffding and Vapnik-Chervonenkis (VC) \cite{abu-mostafa_2012} require that training and test distribution must be the same i.e. they must be generated from the same distribution of the input space. According to Abu-Mostafa et al. 2012: “If the data is (inadvertently) sampled in a biased way, learning will produce a similarly biased outcome.”  If the training data is generated with exclusion of a certain part of the input space, or chemical space, the final model trained with it may not generalize \cite{abu-mostafa_2012} . Conversely, the test data obtained from only a small part of the chemical space, may not suffice as a true proxy to gauge the out-of-sample error. We have explained this in detail in our another work \cite{Athar_2026} . Therefore, a “fair” split of the train/test data, based on the chemical space of the dataset, is extremely crucial. Here, it is important to note the hierarchical structure - based on materials (element dependent), compositions (concentration dependent), and data points (temperature dependent) \cite{Athar_m_j_2025} - of the experimental TE datasets. A train/test split based on the lower levels of the hierarchy (concentration or temperature dependencies) can give misleading test scores as they would only reflect the interpolation ability of the model with respect to concentrations or measurement temperatures. While researchers are cautious in splitting the train/test subsets based on the top hierarchy of the materials \cite{na_chang_2022, parse_n_2024}, to the best of our knowledge there is no reported standard method to fairly split the dataset based on the equivalent representation of chemical space along with the target property for TE materials ($zT$). Tranås et al. (2021) have proposed a principal component analysis (PCA) based distribution of train and test data points \cite{tranas_2021} . However, this study was for a DFT-based non-hierarchical dataset and still lacked consideration for the other possible train/test distributions equivalent in chemical space but different in the ease of model fitting. Both the input space and target property distribution of the train/test data determines the ease of model fitting. However, selecting the train-test split after observing the test data may constitute data snooping.\cite{abu-mostafa_2012} .

The second challenge is related to the feature selection method. Derived from several interdependent intrinsic bulk physical properties ($S$, $\sigma$, and $\kappa$), the TE figure-of-merit depends on the complex interplay of physical, structural, and electronic properties of constituent elements as well as bulk structure. Selecting a relevant and adequate number of those properties as primary features is essential to avoid under-representation of zT in ML-training. Conversely, incorporation of a large number of irrelevant and redundant features may not only increase computation time but also result in multicollinearity-caused overfitting \cite{li_2022} . Therefore, feature selection methods are crucial to check relevancy and redundancy of features. They help decrease the size of the feature space by detecting and removing highly inter-correlated features \cite{barua_2025} . Current feature filtering methods are broadly classified into three categories viz.  filter, wrapper, and embedded methods \cite{nagarajan_babu_2021} . Filter method uses a feature ranking function, e.g. Fisher's score or Pearson Correlation coefficient \cite{gu_2011,deprez_robinson2024} , to choose the best features independent of the ML algorithm. Wrapper methods search the feature space to select the best feature subset, for final use, by testing the performance of each subset using an ML algorithm.  Embedded methods select the best feature subsets on the fly as part of the final training process. Each method offers their own unique advantages and challenges. 
While the filter method is quick and model independent, it ignores relationships between features which are crucial for a complex target property like $zT$. Conversely, though wrapper methods consider feature dependencies, they become expensive with large datasets or large feature spaces.
In embedded methods, lastly, though accounting for feature relationships and being relatively scalable, finding small feature subsets is challenging \cite{biswas_2016} .
A common strategy is to use the filter method as a pre-processing step for ranking features which can then be used with a wrapper method to finalize the cut-off via cross-validation (CV) \cite{guyon_2002, bhattacharjee_2022} . In this method, the minimum number of features to be retained is determined using an iterative redundancy analysis of the top ranked features using the saturation point of their CV score. A critical problem with model-specific-wrapper or embedded-methods stems from hyper-parameter tuning. The model performance of a given ML-algorithm is contingent upon the chosen hyper-parameters. Features selected for one set of hyper-parameters may not be the same as for another set. This, therefore, entails optimization of hyper-parameters at each iteration of the cross-validation process. While this adds another layer of complexity and cost to the already expensive wrapper method, such a brute force approach is both necessary and possible for small datasets, as in the case of TE materials, to improve the generalizability of the model.

In this work, we first propose a principal component analysis (PCA)-based train/test splitting method for hierarchical experimental TE datasets which considers multiple possible splits equivalent in chemical space. Thereafter, an integrated approach for optimization of hyper-parameters along with a $k$-best feature filtering method for four different ML algorithms - Random Forest (RF) \cite{breiman_2001} , XGBoost (XGB) \cite{chen_2016} , Neural Networks (NNs) \cite{kabir_2020} , and Sure Independent Screening-Sparsifying Operator (SISSO) \cite{Ouyang_2018} - is presented. Following the finalization of feature selection and hyper-parameter optimization, an ‘ensemble averaging’ approach for all machine learning techniques is adopted. Lastly, a high-throughput screening (HTS) strategy over a vast and chemically realistic compositional space  is performed to predict top-performing never studied TE half-Heusler materials.

\section{2. Methods}

\subsection{2.1 Composition-based feature vectors of half-Heusler materials}

To numerically represent a $A_{1-x_{A'}}A'_{x_{A'}}B_{1-x_{B'}}B'_{x_{B'}}C_{1-x_{C'}}C'_{x_{C'}}$ half-Heusler material for machine learning, we construct a composition-based feature vector using 19 atomic properties for each of the six elemental components: $A$, $A'$, $B$, $B'$, $C$, and $C'$. The candidate elements for $A$, $B$, and $C$ and their respective dopants $A'$, $B'$, and $C'$ are taken from the list in \cite{anand2019} . The details of the crystallographic structure of the hH used in our work can be found in the same study.

These atomic properties, listed in Table S1 of the Supporting Information (SI), were selected based on their reported importance in previous ML studies for different TE materials in the literature. They include fundamental physical and chemical quantities such as atomic number, atomic radius, atomic weight or valence electrons count. 
Values for these properties are extracted from the Mendeleev Python library \cite{mentel_2021} , normalized by the maximum absolute value to ensure comparability across features, while preserving their relative magnitudes. 
Each element is described by a 19-dimensional vector, and the full representation of a material combines six such vectors, resulting in a 114-dimensional vector that encodes a given composition. When a substituent site is empty, the feature values of the parent site were used for that site. 
To represent doping  of a specific site, we append the three concentration variables $x_{A'}$, $x_{B'}$ and $x_{C'}$, which correspond to the dopant concentrations on the $A$, $B$, and $C$ sites, respectively. These three concentrations, ranging from 0 to 0.5, combined with the 114 atomic features, define a 117-dimensional vector that uniquely represents a composition. 

In this work, we have used the ICGM dataset for half-Heulser TE materials that was curated in our previous work \cite{Athar_m_j_2025}. A given chemical composition can appear at multiple temperatures in the dataset, reflecting the temperature dependence of the thermoelectric properties. 
As a result, the full dataset includes several data points per composition, each associated with a different temperature, which allows for learning temperature-dependent trends across a broad temperature range. Each of these 118-dimensional (117+$T$) input vectors are then used as input for the machine learning model, and the $zT$ values associated with each material and temperature are used as the target property.  This curated dataset comprises 108 distinct materials, 256 unique compositions, and a total of 1521 data points, with temperatures from $289K$ to $1200K$ and $zT$ from $0.0001$ to $1.45$. Our dataset is, to this day, the largest curated dataset, presenting no redundancies, available for half-Heuslers.

\subsection{2.2 Separation of training and test datasets}

To assess the generalizability of our models, that is, their ability to make accurate predictions on previously unseen materials within the half-Heusler chemical space represented by the data, we design a systematic and unbiased approach for splitting our dataset into training and test sets. These splits are based on the distribution of materials in the chemical space defined by the physical properties of the constituent elements of half-Heusler compounds (the 114-dimensional vector described earlier). We aim to construct five folds of our dataset: five separations between training and test sets for which the union of the tests sets forms the full dataset and for which no material is present in two different test sets.
This choice of feature subset leads to the presence of all compositions and temperature points of one material to be exclusively in either the training or the test set of a distribution. 
To construct a meaningful representation of this space, we apply principal component analysis (PCA), a dimensionality reduction technique that projects high-dimensional data into a lower-dimensional space while preserving as much variance as possible. 
PCA identifies orthogonal directions of greatest variance in the data, known as principal components (PCs), which are ordered by their ability to explain the total variance. 
In our case, we retain the first 13 principal components, as determined by a cumulative variance analysis, shown in Figure S1 in the SI, which shows that these 13 PCs account for 95\% of the variance in the dataset. 
This 13-dimensional PCA space becomes the basis of our data splitting. To construct the splits, we first compute the Euclidean distance of each material from the centroid of the dataset in this 13-dimensional PCA space (a 2D illustration is given in Figure 1 (a) where each point corresponds to the 2D projection of a material). 
Materials are then ranked by increasing distance. Each material is consecutively assigned to the test set of the fold corresponding to its rank modulo 5. Hence, materials with ranks 1, 6, 11, 16, \ldots are assigned to the first fold, materials with ranks 2, 7, 12, 17, \ldots to the second fold, and so forth up to the fifth fold. This method ensures that the five test sets are composed of representative and diverse samples in chemical space.

We generate five such train/test folds, ensuring that each fold covers a similar and overlapping region of the chemical space. 
Although for clarity Figure~\ref{fig:tSNE_folds} shows a 2D plot of the distribution of the materials in the five test sets, we use t-SNE\cite{vandermaaten_2008} solely for visualization because it preserves local neighborhoods more faithfully in low dimensions than a linear projection\cite{goodfellow_2016} . This figure shows that materials do not group together for one test set, i.e. the test set of each fold is spread quite homogeneously over the whole space. The actual data split is performed in the full 13-dimensional PCA space computed from the 114-dimensional elemental feature vector; this space is used only to construct the partition, whereas the models are trained on the features selected as described in Section~2.3. The distribution of materials, compositions, and data points in each fold is shown in Figure~S2 in the SI. To assess the sensitivity of the predictive performance to the choice of train/test partition, …we additionally performed 250 independent random material-wise splits, training the RF,
XGB and NN models on the complete set of 118 input features. For each realization, the models
were retrained and the test RMSE was evaluated for both the complete test set and the test set interpolated at 673~K. The resulting RMSE distributions and the convergence of their mean and standard deviation are reported in the SI (Figs.~S9 -- S12 in Section S2).
Tests that withhold entire chemical families or
elements, which probe out-of-distribution extrapolation, are not considered here: for a dataset of the size currently available for thermoelectric materials, each family comprises only a limited number of materials, and withholding one leaves its chemistry unrepresented in training.

\begin{figure}[htbp]
  \centering
  \begin{minipage}[t]{0.58\textwidth}
    \vspace{0pt}
    \centering
    \includegraphics[width=\linewidth]{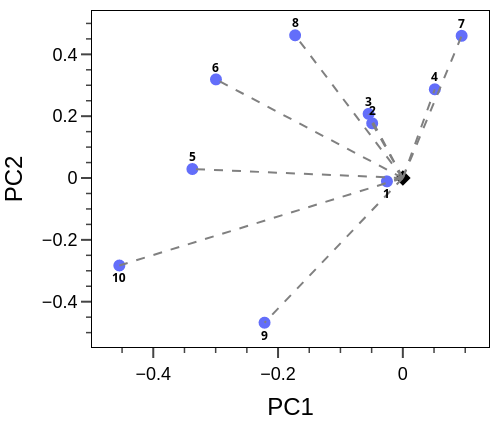}
  \end{minipage}
  \hfill
  \begin{minipage}[t]{0.38\textwidth}
    \vspace{0pt}
    \centering
    \renewcommand{\arraystretch}{1.1}
    \begin{tabular}{@{}cccccc@{}}
      \toprule
      \textbf{Material} & \textbf{F1} & \textbf{F2} & \textbf{F3} & \textbf{F4} & \textbf{F5} \\
      \midrule
1  & \textcolor{red}{Te} & Tr & Tr & Tr & Tr \\
2  & Tr & \textcolor{red}{Te} & Tr & Tr & Tr \\
3  & Tr & Tr & \textcolor{red}{Te} & Tr & Tr \\
4  & Tr & Tr & Tr & \textcolor{red}{Te} & Tr \\
5  & Tr & Tr & Tr & Tr & \textcolor{red}{Te} \\
6  & \textcolor{red}{Te} & Tr & Tr & Tr & Tr \\
7  & Tr & \textcolor{red}{Te} & Tr & Tr & Tr \\
8  & Tr & Tr & \textcolor{red}{Te} & Tr & Tr \\
9  & Tr & Tr & Tr & \textcolor{red}{Te} & Tr \\
10 & Tr & Tr & Tr & Tr & \textcolor{red}{Te} \\
11 & \textcolor{red}{Te} & Tr & Tr & Tr & Tr \\
12 & Tr & \textcolor{red}{Te} & Tr & Tr & Tr \\
13 & Tr & Tr & \textcolor{red}{Te} & Tr & Tr \\
    \end{tabular}
  \end{minipage}
  \caption{a) Illustration in 2D of the ordering of the materials by increasing distance from the centroid of the PCA space. b) Distribution of the materials in the training (Tr) or test (Te) sets for all the folds, based on the order defined in a).}
  \label{fig:pca_construction}
\end{figure}

\begin{figure}[htbp]
    \centering
    \includegraphics[width=1\linewidth]{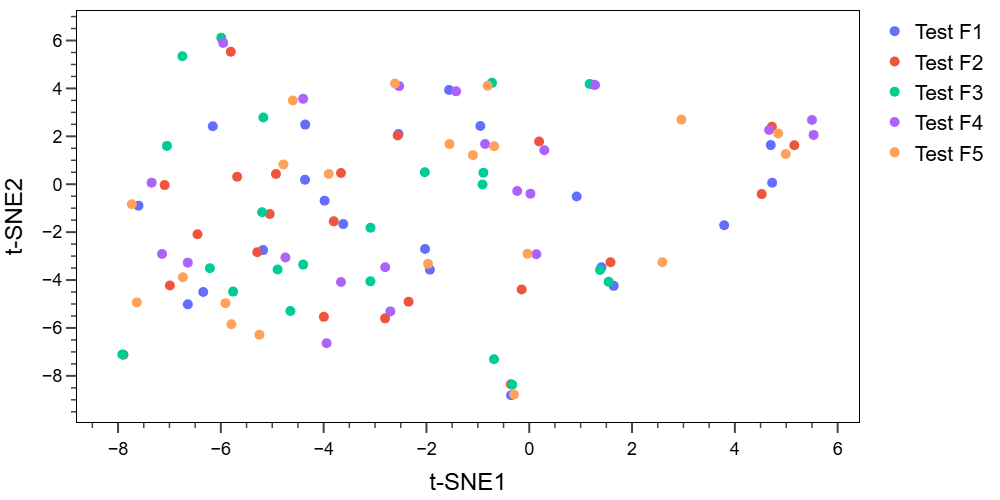}
    \caption{Two-dimensional t-SNE projection of each fold's test set}
    \label{fig:tSNE_folds}
\end{figure}

\subsection{2.3 Feature selection}

\subsubsection{2.3.1 Correlation between elemental features and \boldmath{$zT$}}


For a complex material property such as the TE figure-of-merit $zT$, one might be tempted to use as many elemental features (EFs) as possible to capture their possible relationships with the target property. However, too many input features increase the computational cost and may lead to overfitting, particularly in the presence of multicollinearity, where redundant or highly correlated features can cause the model to learn patterns that do not generalize well to new materials. Conversely, selecting too few features may prevent the model from capturing essential physical relationships, leading to underfitting and reduced predictive performance.

The element-based definition of the 114 site-dependent features implies that the same set of EFs characterizes all compositions and temperatures associated with a given material system. This creates a hierarchical asymmetry in the dataset: while each material may be associated with multiple $zT$ values measured at different temperatures and dopant concentrations, its elemental nature is represented by a unique set of EFs. To calculate statistical correlations between EFs and $zT$, we therefore reduce this hierarchy through successive aggregation. First, for each composition, we retain the maximum $zT$ over all experimentally sampled temperatures. Subsequently, for each material system, we retain the maximum of these composition-level $zT$ values over the available compositions. This ``double maximum'' procedure assigns a single $zT$ value to each material system, thereby establishing a one-to-one mapping between its elemental feature set and its highest experimentally sampled thermoelectric performance.

This choice of aggregation is physically motivated by the objective of identifying elemental characteristics associated with high-performing thermoelectric materials. Temperature and dopant concentration are treated as explicit variables in the dataset rather than as elemental features; consequently, for a given set of constituent elements, including the dopant, the elemental features remain fixed while $zT$ varies with temperature and composition. The maximum aggregation therefore captures the highest experimentally demonstrated $zT$ associated with a given elemental feature set within the sampled temperature and doping space. Importantly, this value represents the highest reported/sampled $zT$ available in the curated dataset and should not be interpreted as an absolute theoretical maximum.

The resulting one-to-one mapping provides a consistent basis for calculating EF--$zT$ correlations while retaining the material-level objective of identifying elemental features associated with high thermoelectric performance. The dataset used for this analysis has also undergone the round robin filtering procedure described previously \cite{Athar_m_j_2025}, which reduces the influence of anomalous measurements and inter-laboratory inconsistencies.

\begin{figure}
    \centering
    \includegraphics[width=1\linewidth]{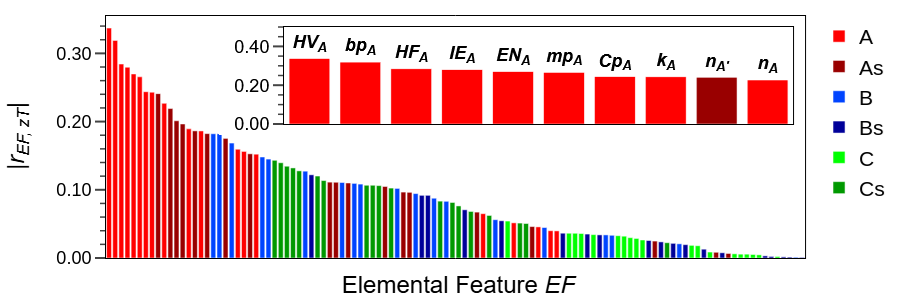}
    \caption{Ranking of the elemental features by absolute Pearson coefficient}
    \label{fig:feature_ordering}
\end{figure}

The Pearson correlation coefficient\cite{pearson_1895} between a given elemental feature $EF$ and the aggregated property $zT$ is then computed as

\begin{equation}
r_{EF,zT} =
\frac{\sum_{i=1}^{n} \left(EF_i - \overline{EF}\right)\left(zT_{i} - \overline{zT}\right)}
{\sqrt{\sum_{i=1}^{n} \left(EF_i - \overline{EF}\right)^2} \, \sqrt{\sum_{i=1}^{n} \left(zT_{i} - \overline{zT}\right)^2}}
\end{equation}

where $\overline{EF}$ and $\overline{zT}$ respectively denote the averages of the EF values and of the $zT$ values across the $n$ materials. By definition, $r_{EF,zT}$ is bounded by $-1$ (perfect anti-correlation) and $1$ (perfect correlation). In our analysis, we rely on the absolute value of the correlation coefficient, since both positive and negative correlations can carry predictive power: even an anti-correlated feature may provide useful information to the models for determining $zT$. We can then sort the features by importance score, as illustrated in Figure~\ref{fig:feature_ordering} . The fold specific ranking of elemental features are shown in Figure S3 in the Supoorting Information.

\subsubsection{2.3.2 Feature elimination}

Following the ranking of the EFs, we proceeded to detect and remove redundant features by constructing Pearson correlation matrices between all feature pairs.
It is important to note that the distributions of the feature values are site dependent. Properties such as the atomic radius ($AR$) or atomic weight ($AW$) exhibit low overall correlation ($r_{AR, AW}=0.49$) for all the elements constituting hH materials regardless of the site. Nevertheless, due to the reduced list of elements authorized per site \cite{anand2019} , the site-specific correlations between features can deviate substantially from this global trend. For example, the correlation coefficient of $AR_C$ and $AW_C$ for the elements possible on site $C$, differs significantly ($r_{AR_C, AW_C} = 0.89$) from the global trend. 
 
\begin{figure}
    \centering
    \includegraphics[width=1\linewidth]{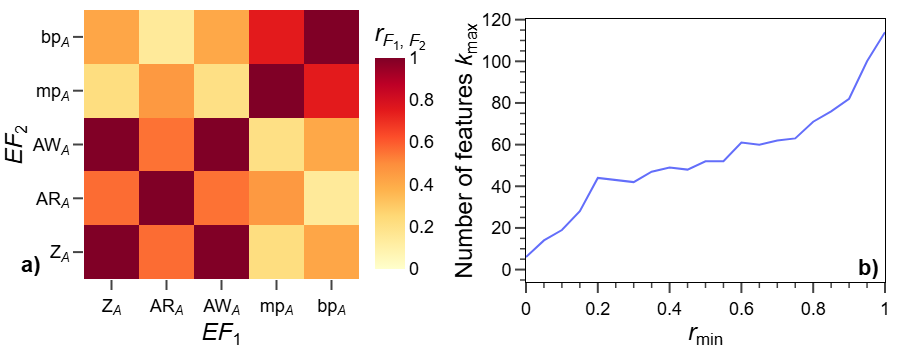}
    \caption{(a) Pearson correlation matrix of five sample elemental features of the A-site elements; (b) Evolution of the number of features $k_\mathrm{max}$ retained as a function of the $r_{\mathrm{min}}$.}
    \label{fig:corr_to_feature_set}
\end{figure}

To account for this, we constructed three separate correlation matrices, one for each site as shown in Figure~\ref{fig:corr_to_feature_set} (a) for the $A$ site  (only five sample features are displayed, the full correlation tables for all sites are available in Figures S4 in the SI).
Because the set of candidate elements at a site remains unchanged regardless of whether substitution occurs, the distributions of possible elemental features at sites $A$ and $A'$ are identical. Consequently, the correlation matrices of their properties coincide. The same reasoning applies to the $B/B'$ and $C/C'$ sites.
To discard highly correlated pairs, we started by constructing a graph by introducing a correlation threshold $r_{\mathrm{min}}$: vertices represent features and edges connect pairs with correlation above $r_{\mathrm{min}}$.  
The Bron–Kerbosch algorithm \cite{bron_1973} was then applied to identify maximal cliques, i.e., fully connected subgraphs that are not contained within larger ones, each corresponding to a group of strongly correlated features.  
From each clique, only the feature showing the highest correlation with $zT$ was retained. 
Repeating this procedure for the six sites yielded a reduced, non-redundant feature list for every chosen value of $r_{\mathrm{min}}$, as reported in Figure~\ref{fig:corr_to_feature_set} (b). 
While the overall number of selected features increases with $r_{\mathrm{min}}$, occasional spikes appear because some thresholds produce different cliques that do not select the same representative feature, whereas higher $r_\mathrm{min}$-produced cliques yield only one feature.

Crucially, the correlation threshold introduced earlier determines how redundancy is handled in the feature set and directly shapes the final model. A high threshold links only nearly identical features, limiting dimensionality reduction and leaving redundancies intact. A low threshold, in contrast, treats even weak correlations as redundant, potentially discarding complementary information. In both cases, the ranking of EFs and the resulting feature subset are affected, which in turn impacts model performance. The correlation threshold must therefore be carefully tuned to strike the right balance between preserving informative diversity and eliminating `redundant` features.

\subsection{2.4 Evaluation metrics}

In this study, we report global predictive performance of ML models comparing $N$ experimental $zT$ values (noted $\zTxp$) and predicted $zT$ values (noted $\zTpr$) using Root Mean Squared Error (RMSE) and Maximum Absolute Error (MaxAE), while also considering Mean Absolute Percentage Error (MAPE) given by the following equations:

\begin{equation}
\mathrm{RMSE} = \sqrt{\frac{1}{N} \sum_{i=1}^{N} \left( \zTpri - \zTxpi \right)^2 }
\end{equation}

\begin{equation}
\mathrm{MAPE} = \frac{100}{N} \sum_{i=1}^{N} \left| \frac{\zTpri - \zTxpi}{\zTxpi} \right|
\end{equation}

\begin{equation}
\mathrm{MaxAE} = \max_{i=1}^{N} \left| {\zTpri - \zTxpi} \right|
\end{equation}

In addition, we define partial metrics to focus on regions of the $zT$ spectrum with particular interest for HTS. 
For example MAPE is computed only on data points with $\zTpr>\cutoffvalue$ for doped compounds. This threshold reflects a practical cutoff for thermoelectric performance, highlights the model’s capacity to correctly predict materials that are experimentally known to be promising and avoids the situation where very small experimental values inflate artificially the MAPE.

\subsection{2.5 The SISSO (Sure Independent Screening with Sparsifying Operator)  method}

The SISSO- Sure Independent Screening (SIS) in combination with Sparsifying Operator (SO) method was employed for descriptor identification from the dataset such that the target property ($zT$) can be expressed as a symbolic linear combination of complex descriptors: 

$\zTpr = \sum_{i=1}^{n} c_id_i + A$\\
where $d_i$ is the $i$th descriptor, $n$ the chosen number of descriptors (the so-called dimension of the regression) and $A$ the intercept in the $n$-dimensional descriptor space.

The method, first, involves the construction of a large number of complex features $\phi^c$ ($10^5$-$10^{12}$ elements) using a combination of algebraic/functional operations recursively on the feature set $\Phi$ (here the feature set $\Phi$ designates the 114 elemental features, the 3 concentrations, and the temperature). The operator set is defined as: \\ 
$\hat H^m \equiv \{I, +,-, \times, /,\exp{} , \log{} , \lvert \rvert, \sqrt{} ,^{-1} ,^{2} ,^{3} , scd\}$[$\phi_1, \phi_2$] - where $scd(\phi) = 1/(\pi(1+\phi^2))$, $\phi_1$ and $\phi_2$ are features in $\Phi$ and the superscript $m$ indicates that the dimensional analysis is performed to retain only meaningful $\phi^c$s.\\ 
This generates candidate features of increasing complexity where the complexity is the maximum number of operations considered. We have chosen a complexity of five. Thereafter, it employs the so-called ‘Sure Independent Screening’ (SIS) - which is a screening process - to select a smaller number of $\phi^c$s, the size of which depends on a user-defined SIS value (in our case 10000) multiplied by the dimension $n$ (in our case $n=2$). Finally, based on the descriptor dimension $n$, SISSO determines the best features from the vast $\phi^c$ feature space by using the sparsifying $l_0$ constraint  \cite{Ouyang_2018} . \\
SISSO calculations become computationally infeasible with 118 input features $\phi$. Therefore, a variable selection method, as proposed by Guo \textit{et al. (2022)} \cite{Guo_2022} , performs iteratively a variable selection and subsequent SISSO calculations for model optimization. The method first obtains a subset $S_a$ from the large pool of input features $\phi$ by random search prioritizing unvisited features (not yet included in  $S_a$) over visited ones. The first SISSO calculation (iteration no. 0) is performed, and the features appearing in the model are transferred from $S_a$ to the initially empty, $S_b$. New random features are added to $S_a$ and the subsequent SISSO calculations are performed with $S = S_a \cup S_b$. Features appearing in the improved SISSO model (better training RMSE) are transferred from  $S_a$ to $S_b$ at each iteration. Finally, the iterations are deemed converged if the RMSE does not improve beyond a user-defined number of iterations, in our case 20. 

\subsection{2.6 High-throughput screening}

\newcommand{\ThtsK}{673\ \mathrm{K}}
\newcommand{\hullnotation}{\Delta E}
\newcommand{\hullcutoffOQMD}{0.15} 
\newcommand{\hullcutoffMP}{0.9}    

\subsection{2.6.1 Thermodynamic and electronic stability filtering}

We first identified candidate $ABC$ parent matrices from the OQMD \cite{saal_2013}and the Materials Project (MP) \cite{jain_2013} databases according 
to our elemental definition of the hH chemical space \cite{anand2019}. For OQMD, entries reported 
in the $F\bar{4}3m$ space group were initially extracted, followed by hH-specific 
elemental filtering. We calibrated database-specific hull-distance cutoffs using 
the 29 experimentally verified parent $ABC$ hH matrices contained in our training 
dataset. For each parent matrix $ABC_i$ and database $D$, we considered the 
minimum hull distance among all corresponding database entries,
\begin{equation}
\Delta E_{\min}(ABC_i,D)
=
\min_{\mathrm{entries}}
\left[
\Delta E_{\mathrm{hull}}(ABC_i,D)
\right],
\end{equation}
and defined
\begin{equation}
\Delta E_{\mathrm{cutoff}}(D)
=
\max_{i=1,\ldots,29}
\left[
\Delta E_{\min}(ABC_i,D)
\right].
\end{equation}
This procedure retains all 29 experimentally verified parent matrices while 
accounting for differences between the two databases. The resulting cutoffs are 
0.15~eV/atom for OQMD and 0.90~eV/atom for MP. A parent matrix was retained if 
its hull distance satisfied
\begin{equation}
\Delta E_{\mathrm{hull}}(ABC)
\leq
\Delta E_{\mathrm{cutoff}}(D)
\end{equation}
in at least one database. Applying this criterion yielded 729 candidate parent 
matrices satisfying the OQMD cutoff and 119 candidate parent matrices satisfying 
the MP cutoff. Combining the two databases and removing duplicate matrices 
retained by both criteria resulted in a total of 631 unique candidate $ABC$ hH 
parent matrices. Parent matrices with no corresponding hull-distance information 
in either database were excluded. The resulting thermodynamic filter is therefore a database-informed screening 
criterion rather than a prediction of experimental synthesizability. In 
particular, it does not explicitly account for finite-temperature stability, 
synthesis kinetics, configurational entropy, or metastable phases. We also did 
not impose separate oxidation-state or competing-phase calculations.

For each of the 631 retained parent matrices, we considered single-, double-, 
and triple-site doping ($A$, $B$, $C$, $AB$, $AC$, $BC$, and $ABC$ substitution 
patterns), drawing candidate dopants from a pool of 25, 14, and 9 elements at 
the $A$, $B$, and $C$ sites, respectively, as proposed by \cite{anand2019}, with 
practical exclusions to preserve experimental feasibility and safety (U, Th, and As). Enumerating the resulting elemental 
substitution combinations across the 631 parent matrices, independent of dopant 
concentration, yields approximately $2.46\times10^{6}$ distinct doped 
compositions. Each of these compositions was subsequently expanded over the 
substitution-concentration grid, sampled with a step of 0.01 from 0 to 0.1 and 
a step of 0.05 from 0.1 to 0.5 at each doped site. Since VEC is a concentration-weighted average of the host and dopant valence-electron counts 
at each site, it varies continuously with $x$ ($x_{A'}$, $x_{B'}$, and $x_{C'}$); we therefore evaluated the 
Slater--Pauling 18-electron criterion,
\begin{equation}
17 < \mathrm{VEC} < 19,
\end{equation}
at each sampled concentration, retaining only compositions satisfying this 
constraint. This criterion serves as an electronic screening constraint and is 
not intended to replace explicit phase-stability calculations. From the 
resulting concentration-resolved compositions satisfying the VEC criterion, the 
highest-performing candidates were identified at the application-relevant 
temperature of 673~K through within-technique and across-technique 
inverse-error-weighted ensembles (see below). The resulting candidates 
therefore satisfy the adopted database-informed thermodynamic and VEC 
constraints but are not guaranteed to be experimentally synthesizable. The purpose of these filters is to reduce the combinatorial doped chemical space to 
a tractable set of computationally prioritized candidates for subsequent 
first-principles and experimental investigation.

\subsubsection{2.6.2 HTS temperature interpolation}
\newcommand{\Txp}{T^\mathrm{xp}}
Because experimental $zT$ is reported at multiple temperatures per composition while models operate at a given temperature, we interpolate each model’s fold-specific test predictions to $\ThtsK$ taken as the working temperature of hHs \cite{Xie_2012} . Concretely, for each fold and each technique, we fit a quadratic surrogate $f$ such as $\zTxp=f(\Txp)$ on the fold’s test compositions (each composition corresponding to a unique $f$) and evaluate models on the interpolated data points at $\ThtsK$, comparing $\zTpr$ and $f(\ThtsK)$. This reduces the number of effective data points used for error estimation (only the interpolated value at $\ThtsK$ per test composition is needed) and aligns the evaluation of a model's predictive performance with the HTS target temperature.

\subsubsection{2.6.3 Within-technique ensemble averaging (5-fold) and across-technique averaging}
\newcommand{\fold}{F}
\newcommand{\efold}{e_{\fold}}
\newcommand{\tech}{t}
\newcommand{\zTtech}{zT_{\tech}^{\mathrm{pr}}}
\newcommand{\zTtechfold}{zT_{\tech, \fold}^{\mathrm{pr}}}

Let $\efold$ denote the RMSE of fold $\fold$ at $\ThtsK$ for a given technique, computed on the interpolated fold test set.
For a fixed technique, the prediction for a candidate composition at $\ThtsK$ is obtained by inverse-error weighting over the five fold models. For a ML technique $t$, we calculate the predicted value as:

\begin{equation}
\zTtech
=
\frac{\displaystyle\sum_{\fold=1}^{5} \frac{\zTtechfold}{\efold}}
     {\displaystyle\sum_{\fold=1}^{5} \frac{1}{\efold}}
\end{equation}

with ${\zTtechfold}$ the prediction from fold $\fold$ and $\efold$ the error of the fold's model at $\ThtsK$. Each technique yields a fold-aggregated prediction $\zTtech$ per candidate composition. \\
To combine techniques, we use a technique-level error $e_\tech$: the RMSE at $\ThtsK$ recombined over $100\%$ of the dataset by merging the test predictions as described before. The final technique-pondered prediction is derived from the same inverse-error scheme at the technique level:

\newcommand{\zTpon}{zT_{\mathrm{pon}}^{\mathrm{pr}}}
\begin{equation}
\zTpon
=
\frac{\displaystyle\sum_{\tech} \frac{\zTtech}{e_\tech}}
     {\displaystyle\sum_{\tech} \frac{1}{e_\tech}}
\end{equation}

For each technique $\tech$ we rank candidates by $\zTtech$ at $\ThtsK$ and retain a top-$N$ list, enforcing one composition per material $(A, A', B, B', C, C')$. 
In parallel, we perform the same procedure across techniques using the technique-pondered predictions, yielding a second shortlist. 
Together, these two lists will serve as inputs in our future work for subsequent DFT calculations and experimental synthesis.

\subsection{2.7 First-principles calculations}
 
First-principles calculations were performed with VASP~\cite{Kresse1996} using the projector augmented-wave method~\cite{Blochl1994,Kresse1999} and the PBE functional~\cite{Perdew1996}, with spin--orbit coupling included self-consistently in a non-collinear treatment. The potentials include 12 (Ce: 5s, 5p, 4f, 5d, 6s), 9 (Co: 3d, 4s; Ir: 5d, 6s) and 5 (Bi: 6s, 6p) valence electrons. No Hubbard correction was applied, since the Ce 4f states lie well above the Fermi level and the band edges are dominated by transition-metal d and Bi 6p states (Section~3.4.2). A plane-wave cutoff of 500~eV and a $25 \times 25 \times 25$ Monkhorst--Pack mesh~\cite{Monkhorst1976} were used. Lattice parameters and atomic positions of the $F\bar{4}3m$ structures were relaxed until residual forces fell below 0.01~eV\,\AA$^{-1}$, with an electronic convergence of $10^{-8}$~eV and Gaussian smearing of 0.01~eV; both compounds relax to non-magnetic ground states. The same settings were used for MgAgBi, MgAuBi and MgAuSb. Transport coefficients were obtained with BoltzTraP2~\cite{Madsen2006,Madsen2018} by Fourier interpolation of the band energies, solving the linearized Boltzmann equation within the constant relaxation time approximation ($\tau = 10^{-14}$~s, a standard value for half-Heusler compounds~\cite{Berland2019}) between 300 and 973~K. All transport calculations were performed at zero net doping ($n = p$), which corresponds to the intrinsic regime for CeCoBi and to the charge-neutral stoichiometric compound for gapless CeIrBi.

\section{3. Results and discussion}

\subsection{3.1 Feature filtering}

\subsubsection{\boldmath{3.1.1 $k$}-best selection}

In this work, each half-Heusler composition is represented by a set of elemental and concentration features that encode its chemical and physical makeup. 
The critical reduction of elemental features (EF) was performed in two successive stages. The first stage applies a correlation-based filtering procedure, a standard technique whose details are provided in the methods section. On top of this established approach, we introduced a second stage that combines feature selection with hyperparameter optimization. While both components exist separately in the literature, their joint application provides a more tailored descriptor set for the prediction of $zT$ in half-Heusler compounds, thereby forming the basis for the subsequent machine-learning results.

\begin{figure}
    \centering
    \includegraphics[width=1\linewidth]{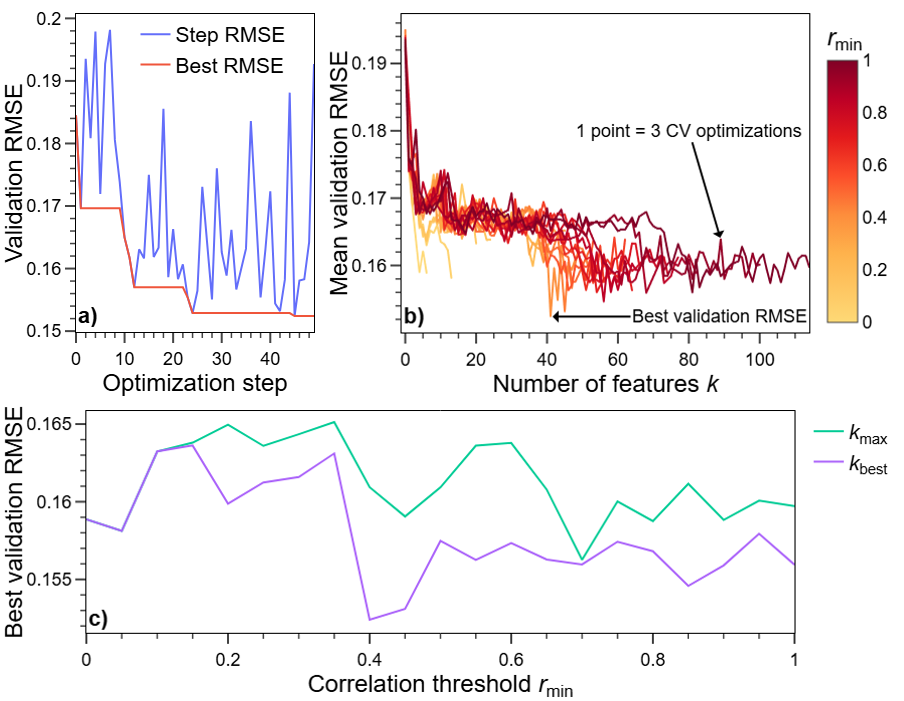}
    \caption{a) Optimization of the XGBoost hyper-parameters with the features set by the pair $(r_{\mathrm{min}}, k) = (0.4, 41)$;  b) $k$-best selection: the pair $(r_{\mathrm{min}}, k) = (0.4, 41)$ leads to a validation RMSE of $0.152$; c) Effect of $k_\mathrm{best}$ over the use of $k_\mathrm{max}$: decrease of the validation RMSE and of the threshold $r_{\mathrm{min}}$ leading to a decreased number of features.}
    \label{fig:kbest}
\end{figure}

Having established the correlation-based filtering of features according to a correlation threshold $r_\mathrm{min}$, we then use a systematic $k$-best procedure to identify the most effective number of elemental features (EFs). This procedure identifies the feature subset that minimizes cross-validation (CV) error. First, the correlation threshold $r_\mathrm{min}$ is varied across a defined interval (twenty evenly spaced values between $-1$ and $1$). For each $r_\mathrm{min}$, a unique list of non-redundant features is constructed using the matrix method described previously and ordered according to their correlation strength with $zT$. As illustrated in Figure~\ref{fig:kbest}(b), the number of features in these lists increases with the threshold, yielding for each threshold an ordered sequence of candidate features. Evaluation of feature subsets is then performed by progressively increasing the subset size. Models are trained using only the first $k$ features in the ordered list, where $k \in \{1,2,\ldots,N\}$ and $N$ is the total number of features retained at the selected $r_\mathrm{min}$. Thus, for each $r_\mathrm{min}$, the $k$-best subset contains the $k$ most $zT$-relevant non-redundant EFs available at that threshold. Each combination of $r_\mathrm{min}$ and $k$ therefore defines a unique feature subset. For each $k$-long feature list, hyperparameter optimization of RF, XGB, and NN (the multilayer-perceptron architecture used for the NN is shown in Figure~S5 in the SI) is performed using Bayesian optimization \cite{bergstra_2011,bergstra_2013}. The corresponding search spaces and the optimized fold-specific hyperparameters are reported in Tables~S3--S5 of the SI. The optimization is performed independently within the 80\% training partition of each of the five PCA-based folds, while the corresponding 20\% test set is kept strictly isolated for the final evaluation. Thus, neither the test data nor any information derived from them is used during feature selection, hyperparameter optimization, or model training.

To ensure unbiased model selection, each 80\% training partition is further divided using a 10-fold CV scheme constructed with the same PCA-based procedure. For each internal fold, the model is trained on 90\% of the 80\% training partition (72\% of the total dataset) and validated on the remaining 10\% (8\% of the total dataset). For a given candidate feature subset and hyperparameter set, the optimization objective is the aggregated RMSE obtained from predictions across the ten internal validation subsets. Thus, every point contributing to this RMSE is evaluated by a model that was not trained on that point. The evolution of this internal-CV RMSE is presented in Figure~\ref{fig:kbest}(a) for an illustrative XGBoost model as an examle.

To mitigate the stochasticity of the Bayesian optimization and ML algorithms, the entire optimization is repeated three times. The mean internal-CV RMSE across the three repetitions is used to determine, for each PCA-based fold, the optimal combination of $(r_\mathrm{min},k)$ and the corresponding hyperparameters. The selected feature subset and hyperparameters are then fixed, and the final model is retrained on the complete 80\% training partition, producing five models for each ML technique, one for each PCA-based fold. Figure~\ref{fig:kbest} (b) illustrates the $k$-best selection for an example XGBoost model. Each curve corresponds to a different value of $r_\mathrm{min}$, and the minimum identifies the combination of $(r_\mathrm{min},k)$ yielding the lowest internal-CV RMSE. For a given threshold, the selected $k_{\mathrm{best}}$ is generally smaller than $k_\mathrm{tot}$, the total number of features retained at that $r_\mathrm{min}$. This reduction in dimensionality, illustrated in Figure~5(c), decreases the number of features entering the model while improving predictive performance. Under the same PCA-based partition, models trained on the full 118-feature input yield test RMSEs of 0.225 (NN), 0.197 (XGB) and 0.213 (RF), compared with 0.164--0.169 for the selected subsets (Table~2), a reduction of 16--27\%. The reduced feature sets also lower the computational cost of the subsequent high-throughput screening.
This nested procedure provides a leakage-controlled strategy for feature selection and model optimization while preserving the independence of the test sets. By rotating through the five PCA-based folds, every data point is ultimately predicted by a model for which it was excluded from feature selection, hyperparameter optimization, and training. The resulting predictions are subsequently combined to construct the test-mode evaluation used to assess model performance under conditions representative of high-throughput screening, where predictions must generalize to previously unseen materials.

\newcommand{\scd}{\operatorname{scd}}

\subsubsection{3.1.2 Feature filtering using iterative variable selection (VS) algorithm for SISSO}

The k-best wrapper approach discussed above is not computationally feasible with the SISSO method, since SISSO relies on the brute search of the high-dimensional feature space. For this method, we used a new algorithm proposed by  Guo et al. (2022)\cite{Guo_2022} that integrates SISSO with iterative variable selection (random search) for optimization of the model with a large number of input features. The details of this method were given in the Methods section. Figure 6 shows the plot of lowest training RMSEs obtained as a function of iterations for a  random search of features. The steep drop in RMSE for all the folds after a few iterations is due to the inclusion of temperature (T) in the feature subset used for training the SISSO models. All the models converge within 60 iterations and the final training RMSEs obtained were in the range of \~0.13 to \~0.16. Table 1 shows the number of occurrences of features in the final SISSO descriptors for all the folds. Temperature, obviously, was the most recurrent feature appearing in all the five folds followed by $x_{A'}$, $HF_{A}$, $HV_{C}$ appearing in three of the folds. The EFs $ea_{A}$, $ea_{A'}$, $ea_{C'}$,$HV_{A}$ and $HF_{C}$ appeared in at least two of the folds whereas the rest of the features were only present in a specific fold. While the $zT$ dependence on substituent concentration is obvious, $x_{B'}$ and $x_{C'}$ were surprisingly missing from all the folds. Therefore, the presence of other features for B and C sites suggest that the identity of the substituent at these sites is, statistically, more influential than its concentration - at least for our dataset.

\begin{figure}
    \centering
    \includegraphics[width=1\linewidth]{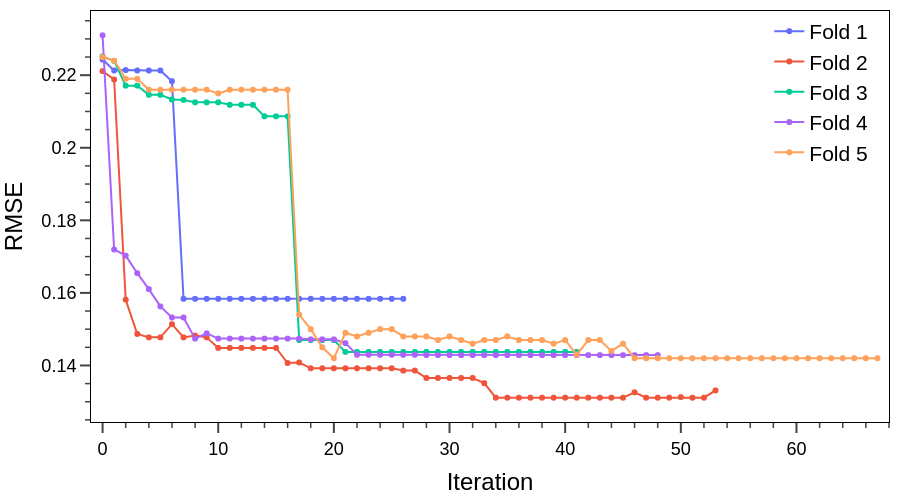}
     \label{Figure 5} \caption{Evolution of the RMSE of the SISSO model during the iterative variable selection process}

\end{figure}

\begin{table}
\centering
\caption{Number of occurrences of elemental features in the final SISSO descriptors across all the folds}
\label{tab:Table 1}
\begin{tabular}{| l | l | l | l |}
\hline
  & \textbf{Elemental Features (EFs)} & \textbf{Number of occurrences} & \textbf{Fold} \\
\hline
1. & T & 5 & F1, F2, F3, F4, F5 \\
\hline
2. & $x_{A'}$ & 3 & F2, F3, F5 \\
\hline
3. & $HV_{A}$ & 3 & F1, F4, F5 \\
\hline
4. & $HF_{C}$ & 3 & F1, F3, F5 \\
\hline
5. & $ea_{A}$ & 2 & F1, F3 \\
\hline
6. & $HF_{A}$ & 2 & F2, F5 \\
\hline
7. & $ea_{A'}$ & 2 & F2, F3 \\
\hline
8. & $HV_{C}$ & 2 & F2, F5 \\
\hline
9. & $ea_{C'}$ & 2 & F1, F2 \\
\hline
10. & $mp_{A}$ & 1 & F5 \\
\hline
11. & $mp_{A'}$ & 1 & F1 \\
\hline
12. & $k_{A'}$ & 1 & F5 \\
\hline
14. & $bp_{B}$ & 1 & F3 \\
\hline
15. & $n_{B}$ & 1 & F4 \\
\hline
16. & $IE_{B}$ & 1 & F2 \\
\hline
17. & $IE_{B'}$ & 1 & F5 \\
\hline
18. & $n_{B'}$ & 1 & F3 \\
\hline
19. & $AR_{C}$ & 1 & F2 \\
\hline
20. & $mp_{C}$ & 1 & F2 \\
\hline
21. & $Z_{C}$ & 1 & F4 \\
\hline
22. & $Z_{C'}$ & 1 & F4 \\
\hline
23. & $Z_{A}$ & 1 & F4 \\
\hline
24. & $Z_{A'}$ & 1 & F4 \\
\hline
25. & $AR_{A'}$ & 1 & F1 \\
\hline

\end{tabular}

\end{table}

\subsection{3.2 Evaluation of ML models}

All ML models are evaluated on a reconstructed test set obtained by combining the predictions from each of the five folds. This allows every data point to be evaluated as unseen data, since its prediction is generated by a model that was not trained on it. The final evaluation is applied consistently across SISSO, NN, XGB, and RF. The parity plots for all the techniques are shown in Figure~7, whereas the fold-wise RMSE values, together with the overall RMSE, are
reported in Table~2. The fold-wise results show that the predictive error varies across the five material-level test folds. This fold-to-fold variation reflects differences in the chemical composition and difficulty of the held-out material subsets, while the overall RMSE provides the aggregate measure of predictive performance across the complete dataset.

As an additional assessment of the sensitivity to the train/test partition, we performed 250 independent random material-wise splits and compared their test RMSE distributions with those obtained for the PCA-based partition using the same 118 input features and model settings
(RMSE = 0.225, 0.197 and 0.213 for NN, XGB and RF, respectively; Figs.~S9--S12). SISSO was not
included in this comparison, as it cannot be applied directly to the full 118-feature space. These values are higher than those in Table~2, which are obtained with the selected feature
subsets (Section~3.1.1). The PCA-based RMSEs lie in the upper part of the random-split distributions (mean $\approx$ 0.18), indicating that the PCA-based partition yields more
demanding test sets than a typical random partition. The random splits also exhibit appreciable variability between individual partitions, demonstrating that the prediction error can depend on the particular material partition. The PCA-based strategy therefore provides a deterministic, demanding, and reproducible benchmark while maintaining representative coverage of the heterogeneous chemical space.

The corresponding distributions of the absolute prediction error, $|\Delta zT| = |\zTpr - \zTxp|$, are provided in Fig.~S6 of the SI. For all four models, the errors remain concentrated at relatively small values, demonstrating consistent predictive performance across the different ML architectures. The absolute error shows a moderate increase toward the higher-$zT$ regime, while the overall error distributions remain comparable across the four models. Importantly, the data is split at the material level rather than at the individual composition level, as discussed in the Introduction. Consequently, all compositions and temperatures associated with a given material are assigned to the same fold, making the test sets more challenging and providing a more stringent assessment of generalization to unseen
materials. The reconstructed-test-set evaluation therefore provides a consistent estimate of
model performance under the material-level generalization scenario relevant to high-throughput screening.

Overall, SISSO and NN achieve almost similar performance, with respective test RMSEs of 0.163 and 0.164. These RMSEs are slightly higher than the best RMSE value of $\sim$0.14 obtained by \cite{Zhong_2023} and are comparable to the $\sim$0.156 achieved by \cite{barua_2024}. However, Zhong et al.~(2023) used in-lab generated experimental data with uniform synthesis conditions and measurements, whereas our dataset is compiled from the literature and therefore incorporates inter-laboratory uncertainty and round-robin errors \cite{Wang_2015}. Additionally, in both of these works, there is ambiguity regarding the hierarchy used for splitting the test and training data, while the RMSE in the present work is calculated on the reconstructed test set containing predictions for the entire dataset. This provides a more stringent assessment of predictive performance and generalizability. To explicitly quantify the incremental contribution of material-dependent features beyond temperature, we also performed a temperature-only ablation study over the same 250 random material-wise partitions. The evolution of the mean RMSE was monitored, and the complete analysis and corresponding results are provided in the SI (Fig.~S13 in Section~S3). For the complete test set, the temperature-only models exhibit higher mean RMSE values ($\sim$0.215--0.23) than the corresponding models using all 118 input features ($\sim$0.18) over the same random partitions, demonstrating that temperature alone does not capture the predictive information provided by the material-dependent descriptors. This difference is particularly evident for the interpolated
test set at 673~K, where all samples have the same temperature. Consequently, a temperature-only
model receives identical input for every material and produces nearly identical predictions, whereas models incorporating the material-dependent features retain substantial material-to-material variation in their predictions. Since temperature is fixed in this test set, the remaining predictive variation necessarily originates from the additional material-dependent input features. These results provide direct evidence that the predictive performance of the proposed models is not simply a consequence of exploiting the temperature dependence of $zT$, but also relies on material-dependent descriptors.

After SISSO and NN, XGBoost showed the best performance with an overall test RMSE of $\sim$0.166, whereas the RF models were the worst overall with an RMSE of 0.169. The good performance of SISSO and XGBoost can be attributed, at least in part, to their ability to control model complexity and identify informative patterns in a relatively small dataset. In SISSO, the sparse descriptor construction promotes a compact representation of the underlying relationships, while XGBoost employs regularization and sequential tree-based learning to control model complexity and capture nonlinear relationships. As for NN, the good performance could potentially be a consequence of its ability to capture intricate and non-linear relationships from the data, as discussed subsequently in the section on SHAP analysis for NN, XGBoost, and RF models, which is particularly relevant for understanding and predicting a complex physical quantity like $zT$ \cite{Bhadeshia_2009}. This performance could further be improved by incorporating regularization such as $l_1$, $l_2$, elastic net, and dropout methods, or their combinations \cite{Farhadi_2022}. Given that NNs are ``data-glutton'' \cite{dou_2023}, combining ensemble learning strategies with NNs (called as NN ensembles) may prove to be even more
promising due to their ability to handle small structured (preprocessed) data more efficiently \cite{kondratyuk_2020}. The maximum absolute error (MaxAE) provides a complementary measure of the largest individual prediction deviation. The overall MaxAE values are 0.672, 0.724, 0.805, and 0.815 for SISSO, NN, XGB, and RF, respectively, with the corresponding worst-case errors occurring in folds 1, 1, 5, and 2. Although the tree-based models exhibit the largest MaxAE, the relatively close overall RMSE values indicate that these isolated large deviations do not dominate the aggregate
predictive performance. Additional performance metrics, including fold-wise and overall $R^2$ and MAE values, are provided in Table~S6 in the SI.

A systematic trend present in all models is the overprediction for $\zTxp < 0.2$ and underprediction for $\zTxp > 0.2$, often producing a cluster of similar $\zTpr$ values even when $\zTxp$ covers a wider range. The phenomenon was more pronounced for non-SISSO models, showing a large number of outliers. Such datapoints were also visible in the parity plots of \cite{jia_2024, parse_n_2024, barua_2024} but were presumably treated as dataset outliers and not discussed further. Since our data is thoroughly curated, this cannot be attributed to experimental/measurement/data extraction errors but primarily reflects the sensitivity of the models to the data. Further investigation into the raw data revealed that this behavior was mostly limited to compounds with very small experimental $zT$ values in their pristine form, and showing a multifold increase in $\zTxp$ upon doping. For example, Zhu et al.~(2019) reported a dramatic increase in $\zTxp$ of TaFeSb, from 0.0176 at 970~K, to 1.39 for A-site doped
Ta$_{0.84}$Ti$_{0.16}$FeSb at the same temperature \cite{zhu_2019}. A similar situation exists for other materials such as TiCoSb \cite{zhou_2007}, NbFeSb \cite{mohamed_2020}, and ZrCoBi \cite{zhu_2018}. On the other hand, there were several compounds, such as TiNiSn \cite{kim_2007} and ZrNiSn \cite{gong_2019}, with reasonable $\zTxp$ values in pristine form and only showing a small increase with doping. While physically there is no restriction on the magnitude of doping-induced $zT$ improvements, this represents a significant challenge for ML models to incorporate a balance between these two types of materials, especially without synthesis information. This results in $\zTpr$ overpredictions for such pure compounds whereas their doped counterparts give higher but underpredicted $\zTpr$ values. When excluding the pure compounds, the MAPE improves noticeably as shown in the partial MAPE analysis in Figure~S7 in the SI. This is justified in the context of high-throughput screening, where the focus lies on doped compounds. Regardless, although our models compress the $zT$ range, underpredicting the highest values, they achieve stable global accuracy, with the obtained test RMSEs corresponding to approximately 11\%--12\% of the maximum experimental $zT$ value (1.45).

\begin{table}[t]
\centering
\caption{Fold-wise and overall predictive performance metrics of the ML techniques evaluated on the recombined test dataset.}
\label{tab:performance}
\begin{tabular}{llcccc}
\hline
Metric & Fold & SISSO & NN & XGB & RF \\
\hline
RMSE & 1 & 0.158 & 0.149 & 0.132 & 0.154 \\
      & 2 & 0.170 & 0.198 & 0.194 & 0.193 \\
      & 3 & 0.156 & 0.133 & 0.146 & 0.145 \\
      & 4 & 0.176 & 0.142 & 0.168 & 0.175 \\
      & 5 & 0.153 & 0.183 & 0.181 & 0.174 \\
      & Overall & 0.163 & 0.164 & 0.166 & 0.169 \\
\hline
MAPE & $zT_{\mathrm{xp}}$ (doped) $>0.2$ 
      & 30.9\% & 26.6\% & 28.6\% & 30.7\% \\
\hline
MaxAE & Overall 
      & 0.672 (Fold 1) & 0.724 (Fold 1) & 0.805 (Fold 5) & 0.815 (Fold 2) \\
\hline
\end{tabular}
\end{table}

\begin{figure}
    \centering
    \includegraphics[width=1\linewidth]{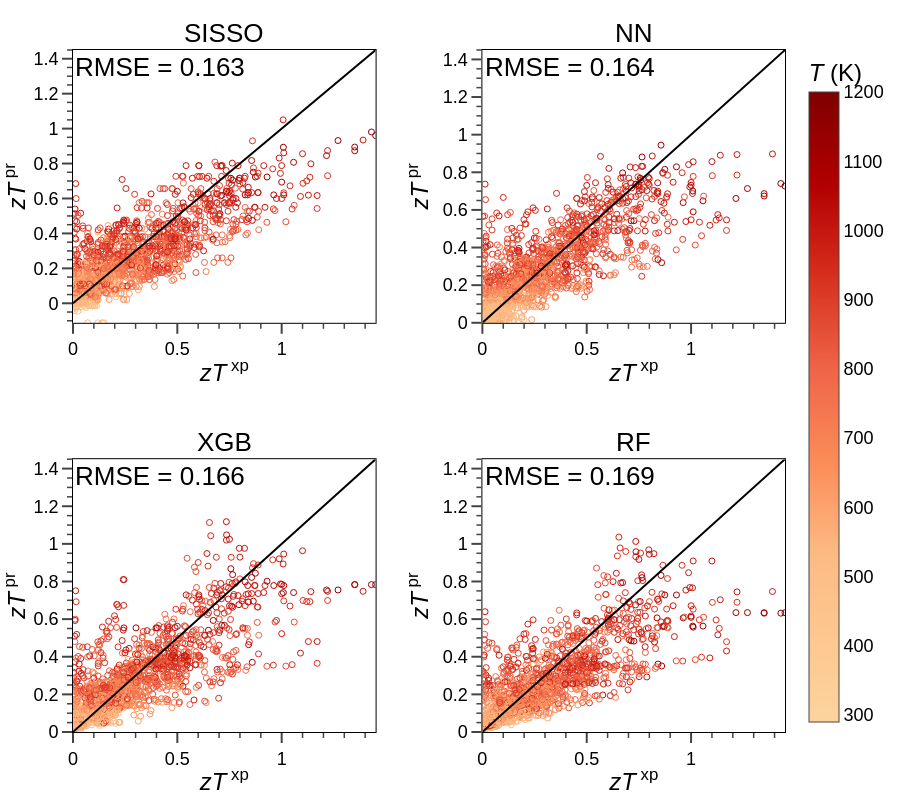}
\caption{Parity plots of predicted $zT$ ($\zTpr$) vs experimental $zT$ ($\zTxp$) for each ML technique.}
\label{fig:parity}
\end{figure}

\begin{table}[htbp]
\centering
\caption{Comparison of predicted and literature-reported $zT$ at the optimal 
dopant concentration and temperature for ten previously published half-Heusler 
systems, one representative composition per cross-validation fold. Predictions 
are evaluated at the exact (composition, $T$) pair reported as optimal in the 
cited source; each system was held out of training in the fold indicated.}
\label{tab:zt_recovery}
\begin{tabular}{lccccccc}
\hline
Material & $T$ (K) & $\zTxp$ [Ref.] & SISSO & NN & XGB & RF & Fold \\
\hline
Nb$_{0.88}$Hf$_{0.12}$FeSb$_{0.88}$Sn$_{0.12}$ & 973  & 0.822 \cite{mohamed_2020} & 0.716 & 0.703 & 0.976 & 0.948 & 1 \\
ZrNiSn                                          & 784  & 0.478 \cite{gong_2019}   & 0.343 & 0.349 & 0.245 & 0.267 & 1 \\
TiNiSn                                          & 753  & 0.376 \cite{kim_2007}    & 0.508 & 0.485 & 0.530 & 0.568 & 2 \\
Nb$_{0.88}$Hf$_{0.12}$FeSb                      & 1200 & 1.450 \cite{fu_2015}     & 0.960 & 0.726 & 0.783 & 0.634 & 2 \\
Ta$_{0.84}$Ti$_{0.16}$FeSb                      & 970  & 1.390 \cite{zhu_2019}    & 0.934 & 0.898 & 0.748 & 0.747 & 3 \\
Hf$_{0.6}$Zr$_{0.4}$CoSb$_{0.8}$Sn$_{0.2}$      & 1123 & 0.856 \cite{hu_2018}     & 0.817 & 0.944 & 0.822 & 0.832 & 3 \\
ZrCo$_{0.9}$Ni$_{0.1}$Bi$_{0.85}$Sb$_{0.15}$    & 973  & 1.007 \cite{zhu_2019b}   & 1.050 & 0.687 & 0.893 & 0.766 & 4 \\
Hf$_{0.65}$Zr$_{0.35}$NiSn$_{0.98}$Sb$_{0.02}$  & 873  & 0.477 \cite{xie_2013}    & 0.529 & 0.650 & 0.583 & 0.527 & 4 \\
Zr$_{0.5}$Hf$_{0.5}$CoSb$_{0.98}$Sn$_{0.02}$      & 1123 & 0.930 \cite{hu_2018}     & 0.723 & 0.829 & 0.802 & 0.728 & 5 \\
Zr$_{0.5}$Hf$_{0.5}$Co$_{0.95}$Ni$_{0.05}$Sb    & 1074 & 1.000 \cite{he_2017}     & 0.694 & 0.731 & 0.785 & 0.609 & 5 \\
\hline
\end{tabular}
\end{table}

Furthermore, to directly assess whether the models can recover established, well-characterized  high-$zT$ half-Heusler systems at their literature-reported optimal doping level 
and temperature, we evaluated model predictions at the exact (composition, $T$)  pair identified as optimal in the original source, for ten representative systems  spanning all five cross-validation folds (Table~\ref{tab:zt_recovery}); since fold assignment follows our material-level split, every reported value is a genuinely  unseen prediction. The set spans established n-type (TiNiSn~\cite{kim_2007}, ZrNiSn~\cite{gong_2019}, NbFeSb~\cite{fu_2015,mohamed_2020}) and p-type (CoSb-based~\cite{hu_2018}, CoBi-based~\cite{zhu_2019b}, Sb-doped NiSn~\cite{xie_2013}) half-Heusler families, including the Zr/Hf-NiSn-Sb-based 
systems widely regarded in the literature for their thermoelectric performance. Across the ten systems, SISSO shows the closest agreement with experiment (MAE = 0.20), followed by XGBoost (0.24), NN (0.25) and RF (0.29). Consistent with the compression of the $zT$ range discussed above, the two highest-$zT$, heavily doped FeSb-based systems (Nb$_{0.88}$Hf$_{0.12}$FeSb and Ta$_{0.84}$Ti$_{0.16}$FeSb) are underpredicted by 0.46--0.82 in $zT$ across all four models, while two of the lowest-$zT$ systems, TiNiSn and Hf$_{0.65}$Zr$_{0.35}$NiSn$_{0.98}$Sb$_{0.02}$, are overpredicted by 0.05--0.19; ZrNiSn, in contrast, is underpredicted by 0.13--0.23. The median absolute errors of SISSO, NN and XGBoost (0.13--0.15) lie below their overall test RMSE, whereas that of RF (0.21) exceeds it. Within
0.33, approximately twice the overall test RMSE, SISSO, NN and XGBoost each predict eight of the ten systems, the exceptions being the two FeSb-based systems, and RF predicts seven. Fold-resolved parity plots for the complete held-out data of each system, spanning all measured dopant concentrations and temperatures rather than only the tabulated optimum (four systems are represented by a single composition), are provided in Fig.~S8 of the SI, with dashed lines marking the experimental and model-predicted $zT$ maxima. Using these data, we tested whether the models locate the optima, comparing compositions at the highest temperature measured for all compositions of a series. We consider a predicted optimum near-optimal if its experimental $zT$ lies within 10\% of the series maximum, a difference within the typical inter-laboratory uncertainty of $zT$ measurements \cite{Wang_2015}. In the six series containing several dopant concentrations, the composition predicted by SISSO is near-optimal in all four series with varying A-site concentration, whereas SISSO cannot discriminate B- or C-site concentrations, which are absent from its descriptors. For NN, XGBoost and RF, the predicted optimum is near-optimal in four, four and five of the six series, respectively, and reaches on average 91--94\% of the experimental maximum $zT$. The temperature of maximum $zT$ is reproduced for 8--9 of the 10 optimal compositions, the exceptions being the systems whose $zT$ passes through a maximum within the measured range. Thus, although the FeSb-based systems are underpredicted in magnitude, SISSO, XGBoost and RF identify near-optimal compositions in both series, whereas NN selects a composition reaching only 64\% of the maximum for Nb$_{1-x}$Hf$_{x}$FeSb. This underprediction therefore reflects the multifold doping-induced increase of $zT$ in these systems rather than a failure to identify their optima.

\subsection{3.3 Interpretation of ML models}

\subsubsection{3.3.1 SISSO Descriptors Interpretation}

Table 3 depicts the final SISSO descriptors of the form $zT$ = $c_{1}$$d_{1}$ + $c_{2}$ $d_{2}$ + A across all the folds -where $d_{1}$ and $d{2}$ are one- and two- dimensional descriptors, $c_{1}$ and $c_{2}$ their respective coefficients, and A the intercept. This represents a symbolic linear expression of $zT$ with respect to individual descriptors but $zT$ is highly non-linear with respect to the elemental features. Apart from F3, the $zT$ vs T relationship in all the folds is linear. From a physical perspective this is not true, since $zT$ values pass through a maximum before declining due to several physical phenomena - such as bipolar conduction, increased electronic thermal conductivity, etc. However, the experimental temperature data of $zT$ is usually limited to peak $zT$ values in the literature and by extension in our dataset. Hence, the model is unable to learn the declining parabolic trend beyond \textit{zT\textsubscript{max}} temperatures. A surprising observation is the absence of $B$ or $B^\prime$- site EFs in F1. Physically, the importance of the $B$-site in hH cannot be underestimated. However, statistically, at least for F1, this means that the influence of $A$- and $C$- sites overwhelms the contribution of B- site EFs to the $zT$. This can clearly be observed from the ranking of the elemental features by the Pearson coefficient, shown in Figure S3 in the SI where $A$-site EFs show a dominant correlation to $zT$ followed by $C$-site EFs (for F1). A significant advantage of ensemble averaging is the inclusion of all local trends in the ultimate average model. Therefore, by virtue of descriptors from other folds representing $B$ or $B^\prime$ sites, their influence is not completely ignored in the final HTS predictions.

For the physical interpretation of descriptors, it is important to avoid an apophenic approach since several terms appearing may just be a consequence of the statistical artifact of the specific fold, or the dataset (as we saw with $x_{B^\prime}$ and $x_{C^\prime}$), and may not necessarily reflect a global trend across hH materials, or other TE materials for that matter. SISSO involves a dimensional analysis to retain only meaningful correlations between target property and combinations of EFs (e.g., no unphysical items such as size + energy or size + $size^{2}$). Therefore, the resulting descriptors are supposed to be "physically meaningful" \cite{Ouyang_2018} . However, there are several EFs which are identical in dimensions (units) but mostly unrelated to each other - such as $HF_{C}$ and $e_{A}$ appearing as $|HF_{C} - ea_{A}|$ in F3 ($d_{2}$). Additionally, the complexity of the descriptors also limits an exhaustive interpretation. Therefore, we must restrict our analysis to only physically intuitive and/or most consistent terms in the descriptors.

The presence of $\scd\!\left(\dfrac{\mathrm{x}_{A^{\prime}}}{\mathrm{ea}_{A^{\prime}}}\right)$, $\sqrt[3]{\dfrac{\mathrm{x}_{A^{\prime}}}{\,\mathrm{ea}_{A^{\prime}}\,}}$, and $\exp\!\left(\dfrac{\mathrm{x}_{A^{\prime}}}{\,\mathrm{ea}_{A^{\prime}}}\right)$ 

(with negative coefficients) in F2 ($d_{2}$), F3($d_{1}$), and F5 ($d_{2}$), respectively, is interesting, especially due the coupling of $x_{A'}$ to $ea_{A'}$ and $k_{A'}$. The terms $\scd$ and $\sqrt[3].$ represent conflicting behavior at high values of $x_{A'}$/$ea_{A'}$. For the $x_{A'}$, a common behavior is its strong non-linear effect on the $zT$ suggesting that the increase in $zT$ will be more prominent for the initial increase in substituent concentrations. This effect can be explained by the fact that a small dopant concentration can result in the formation of impurity bands within the band gap \cite{Zhong_2023, Heremans_2008} . For $ea_{A'}$, the $\scd$ and $\sqrt[3].$ functions require higher and lower values of $ea_{A'}$, respectively, making interpretation difficult. However, we see a more consisting trend of the positive effect of electron affinity on $zT$, as evidenced by the term $ea_{A}$ in F1 and F3. Physically, an atom with a smaller electron affinity can completely lose its valence electrons and serve as a Coulomb active scattering center for charge carriers and, thus, reduce charge mobility. Conversely, an atom with higher electron affinity should (A-site in this case) still attract part of its  negative charge and reduce Coulomb scattering for charge carriers. This can improve carrier mobility, and therefore electrical conductivity, leading to a higher $zT$ \cite{zhu_2021} . While it is tempting to attribute a small $k_{A'}$ to a smaller bulk thermal conductivity end hence to a higher $zT$, the bulk thermal conductivity of a material is not simply a direct sum of the thermal conductivities of its constituent elements. It's rather a complex function of the material's composition, microstructure, and porosity \cite{kim_2017, eivari_2021} .

 The term $|Z_{A} - Z_{A'}|$ in F4 ($d_{1}$) indicates that $zT$ can be maximized with heavier atoms at $A$-site and lighter atoms at the $A^{'}$-site or vice versa. The disparity of weights between $A$ and $A^{'}$ atoms can maximize the point defect scattering, which can result in a decrease of the thermal conductivity and hence an increase of $zT$ \cite{Gurunathan_2020} .The role of alloying, particularly on the A-site, in suppressing the thermal conductivity of hH materials has been consistently reported in the literature \cite{quinn_bos_2021} . It is also important to recall that $HV_{A}$ and $HF_{C}$ have the highest occurrence number in all the descriptors, and, therefore, their influence cannot be ignored. We consistently observe a requirement of large $HV_{A}$ in F1 ($d_{1}$), F4 ($d_{1}$), and F5 ($d_{1}$) descriptors. This can be attributed to the ability of elements at the $A$-site to form stronger bonds with neighboring atoms and, thereby, reduce disorder and maintain higher electrical conductivity at high temperatures \cite{Zhong_2023, ando_1999, Cui_2017} . On the other hand, we observe a requirement of low $HF_{C}$/$HV_{A}$, in F1 ($d_{2}$), and $|HF_A-HV_C$, in F2 ($d_{1}$) and F5 ($d_{1}$), respectively. When coupled together, these observations demand higher $HF_{A}$ and $HV_{A}$ along with a lower $HF_{C}$ and $HV_{C}$. In TE materials, while strong bonds can sustain good electrical conductivity and stability at elevated temperatures, weaker bonds can reduce lattice thermal conductivity, and increase $zT$ \cite{Liang_2022} . This, therefore, requires a fragile balance of bond strengths, and our descriptors implicitly quantify the effect of this balance on $zT$.  To sum up, while we could identify and interpret physically intuitive and consistent trends in the SISSO models, the complexity of SISSO descriptors and uncertainty associated with statistical artifact of the dataset restricted an exhaustive interpretation. Yet, this limited physical interpretability of SISSO is better than other ‘black-box’ algorithms where model analysis is only restricted to feature importance.

\begin{table}[h!]
\centering
\caption{Final SISSO descriptors of the form $\zTpr = c_1d_1+c_2d_2+A$ across all the folds}
\small
\begin{tabularx}{\textwidth}{|c|p{3.8cm}|c|X|}
\hline
\textbf{Fold} & \textbf{Parameters} & \textbf{RMSE (Train/test)} & \textbf{Descriptors} \\
\hline

F1 &
\makecell[l]{%
$ c_1 = 0.962$\\
$ c_2 = -0.001$\\
$ A = 0.141$
} &
0.158/0.158 &
\makecell[l]{%
$ d_1 = T \times {AR}_{A^{\prime}} \times {HV}_{A} \times
\exp\!\left(\dfrac{{HF}_{C}}{{mp}_{A^{\prime}}}\right)$\\[6pt]
$ d_2 = \dfrac{T}{{ea}_{A}\times\displaystyle\left(\dfrac{\mathrm{HF}_{C}}{{HV}_{A}} - \scd\!\bigl({ea}_{C^{\prime}}\bigr)\right)}$
} \\
\hline

F2 &
\makecell[l]{%
$ c_1 = 5.04$\\
$ c_2 = 1.13$\\
$ A = 0.026$
} &
0.131/0.170 &
\makecell[l]{%
$ d_1 = T \times {AR}_{C} \times {ea}_{C^{\prime}}\,\times \sqrt[3]{\lvert {HF}_{A} - {HF}_{C}\rvert} $\\[6pt]
$ d_2 = \dfrac{{mp}_{C}\times \scd\!\left(\dfrac{{x}_{A^{\prime}}}{{ea}_{A^{\prime}}}\right)}
{{HF}_{A}-{IE}_{B}}$
} \\
\hline

F3 &
\makecell[l]{%
$ c_1 = 0.104$\\
$ c_2 = 0.314$\\
$ A = 0.011$
} &
0.144/0.156 &
\makecell[l]{%
$ d_1 = \left(\dfrac{T}{{n}_{B^{\prime}}}\right)^{3} \times \sqrt[3]{\dfrac{{x}_{A^{\prime}}}{\,{ea}_{A^{\prime}}\,}}$\\[6pt]
$d_2 = \dfrac{1}{\exp\!\left(\dfrac{{bp}_{B}}{T}\right)\,\times\sqrt[3]{\lvert {HF}_{C} - ea}_{A}\rvert}$
} \\
\hline

F4 &
\makecell[l]{%
$ c_1 = 0.398$\\
$ c_2 = -4.43$\\
$ A = -0.132$
} &
0.143/0.176 &
\makecell[l]{%
$ d_1 = \dfrac{T \times {HV}_{A} \times \bigl({Z}_{C^{\prime}} + \lvert {Z}_{A} - {Z}_{A^{\prime}}\rvert\bigr)}
{{bp}_{C}}$\\[6pt]
$ d_2 = \dfrac{T \times \lvert {Z}_{C} - {n}_{B}\rvert}{\,{Z}_{C} \big/ \log\!\bigl({Z}_{C^{\prime}}\bigr)}$
} \\
\hline

F5 &
\makecell[l]{%
$ c_1 = 1.96$\\
$ c_2 = -0.942$\\
$ A = 0.05$
} &
0.142/0.153 &
\makecell[l]{%
$ d_1 = \dfrac{T \times {HV}_{A} \times \sqrt{\lvert {HF}_{A} - {HV}_{C}\rvert}}{{mp}_{A}}$\\[6pt]
$d_2 = \dfrac{\bigl\lvert {HF}_{C} - \bigl({IE}_{B^{\prime}} - {HV}_{A}\bigr)\bigr\rvert}
{\exp\!\left(\dfrac{{x}_{A^{\prime}}}{{k}_{A^{\prime}}}\right)}
$} \\
\hline

\end{tabularx}
\end{table}

\subsubsection{3.3.2 SHAP analysis for NN, XGBoost, and RF Models}

The SHAP (SHapley Additive exPlanations) \cite{shap} violin plots for the model-dependent feature importance is shown in figure 8 for the RF, XGB, and NN models. In these plots, the features are ranked from top to bottom (Y-axis) based on their global importance. The top feature is the strongest "driver" of the model's $zT$ predictions. The SHAP values (X-axis) show how much a feature pushes the prediction away from the average (origin) with positive and negative values corresponding to positive and negative correlation to $zT$, respectively. The wide section of the violin shape indicates a large number of data points with similar SHAP values. The color reflects the magnitude of the features (red for high, blue for low). In combination with SHAP values, directional trends can be deduced: blue points on the left/red points on the right, indicate a positive correlation between an increase/decrease of the feature with an increase/decrease of $zT$. Similary to the interpretation of the SISSO descriptors in the previous section, we will limit our analysis to the most consistent or important trends. 

\begin{figure}
    \centering
    \includegraphics[width=1\linewidth]{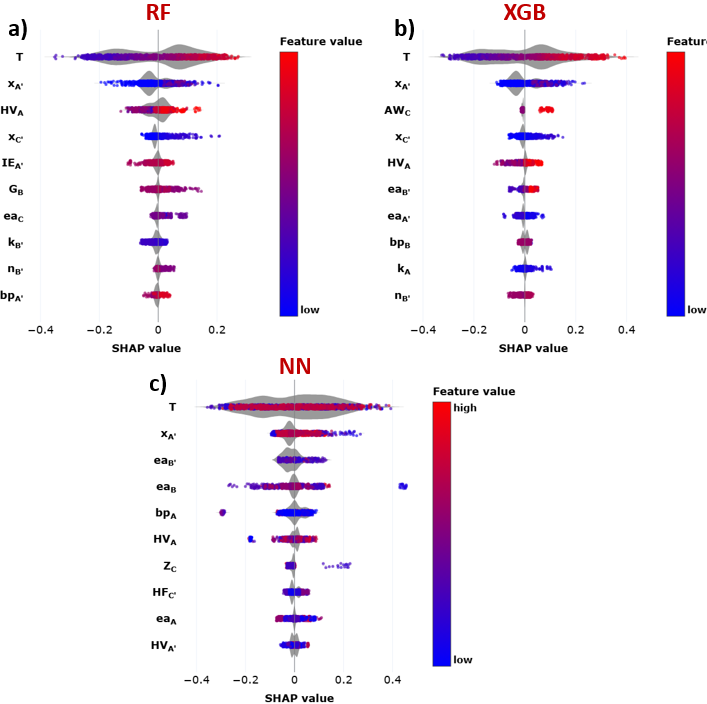}
\caption{Comparative SHAP global analysis for the test sets for a) RF, b} XGB, and c) NN.
\label{fig:SHAP}
\end{figure}

Temperature (T), being both a physical constant in the $zT$ formula and a dynamic environmental variable that changes the material's TE properties, is the most important feature in all three models. For RF and XGB, the distinct segregation of low-value (blue) points on the negative SHAP side and high value (red) points on the positive side suggests a monotonic positive correlation with $zT$. Interestingly, for NN, the red and blue points are mixed on both sides, which indicates that the feature has a complex, non-monotonic relationship with zT— governed by interactions with other features rather than acting as an independent driver. 

$x_{A'}$ (A-site dopant concentration) is the next omnipresent feature across all the models. For $x_{A'}$, however, the SHAP distribution reveals a more nuanced behavior for all models. The presence of blue points on both sides of the zero-baseline suggests that the influence of $x_{A'}$ on $zT$ is context-dependent, governed by feature interactions -similar to what we observe for T in NN models. Nevertheless, the greater density (indicated by the vertical width of the violin) on the negative side of the x-axis demonstrates that, when averaged across the entire dataset, low $x_{A'}$ values more frequently suppress $zT$. This confirms that while a low $A$-site dopant concentration is statistically detrimental to high $zT$, it is a probabilistic trend that can be modulated by the specific material environment. Indeed, the presence of blue points on the extreme right for all the plots proves that for some materials exceptionally high $zT$ with small dopant concentration (at the A-site) can be achieved - a fact corroborated by SISSO descriptors as well as experimentally established for TaFeSb \cite{zhu_2019} , TiCoSb \cite{zhou_2007} , NbFeSb \cite{mohamed_2020} , and ZrCoBi \cite{zhu_2018} as discussed before.

From the model-agnostic ranking of the top 10 EFs in Figure 3, only $HV_A$ (heat of vaporization for A-site) appears in all three SHAP plots—consistent with the trend in SISSO descriptors. Similar to our observation with $x_{A'}$, we see a very subtle SHAP distribution with presence of red points on both the sides of the baseline for RF and XGB. Yet again, statistically, considering the greater density of data points on the positive side for these plots, $HV_A$’s correlation to $zT$ is largely positive for RF and XGB.  NN shows a more subtle, mixed trend with presence of blue and red points on either side. This result, on the one hand, unanimously establishes the importance of $HV_A$ in both model-agnostic and model-dependent frameworks. On the other hand, the fact that no EFs (other than $HV_A$) are simultaneously present in all the plots emphasizes the importance of inter-correlation between EFs rather than their individual importance for $zT$ prediction. This demonstrates that each technique handles these variables differently, with the NN capturing more complex, non-linear relationships.

\subsection{3.4 High-throughput screening (HTS) and validation}

\subsubsection{3.4.1 High-throughput screening of undoped Compositional Space}

Determining the phase stability of a hypothetical compound from \textit{ab initio} calculations requires screening all possible crystal structures for a given composition to identify the phase with the lowest enthalpy of formation. This task becomes prohibitively expensive for high-throughput \textit{ab initio} calculations across vast compositional spaces, particularly given the limited accuracy of density functional theory (DFT) in predicting the stability of half-Heusler (hH) compounds \cite{Legrain_2018, bhattacharya_madsen_2016}. In such cases, machine learning (ML) can facilitate the prediction of the probability that a given chemical composition forms a particular crystal structure. Legrain \textit{et al.}~(2018) \cite{Legrain_2018} used a random forest (RF) classification model to screen 71,178 equiatomic ternary (ABC) compositions for stability in the half-Heusler structure. Of these, 481 compositions were predicted to have a probability greater than 0.5 of being stable in the hH structure, several of which have since been reported experimentally. In the present work, we used this list and adopted a more stringent probability threshold of $>0.6$ to evaluate our ML models. It should be emphasized that the objective here is not to perform an exhaustive high-throughput search (HTS) for hH compounds, but rather to assess the generalizability of our models for previously unexplored hypothetical compositions. Successful predictions would, in turn, provide confidence for a more comprehensive brute-force exploration of the doped compositional space. The top four candidate hH compositions from Legrain \textit{et al.}~\cite{Legrain_2018}, ranked according to the highest predicted $zT$ values at 673~K by each ML model, are listed in Table~5.

From Table~5, a total of 12 unique hH compositions are identified, of which 10 are Bi-based. Notably, approximately 40\% of the stable hH compositions predicted by Legrain \textit{et al.} \cite{Legrain_2018}are Bi-based, followed by Sb-based ($\sim$22\%) and Pb-based ($\sim$14\%) compounds. Although this distribution partly reflects the Bi-rich composition of the candidate pool, the disproportionate representation of Bi-containing compounds also underscores their importance as promising thermoelectric (TE) half-Heusler materials, as demonstrated in previous studies \cite{Pati_2024, Vikram_2017, Panday_2025}. With the exception of LuPdBi, which has demonstrated an impressive room-temperature power factor of $10$~mW~m$^{-1}$~K$^{-2}$ \cite{Mukhopadhyay_2017}, none of these hH compounds had been experimentally reported at the time of our original screening. Likewise, with the exception of CeCoBi and CeIrBi, which are investigated in the present work, subsequent first-principles studies have since been reported for the other seven non-Mg candidates, including NbFeBi, TaFeBi, HfCoBi, HfIrBi, NbOsBi, and LuAuPb. Among these, NbFeBi \cite{Nan_2025}, TaFeBi \cite{Panday_2025}, and HfCoBi \cite{Yang_2022} have demonstrated outstanding thermoelectric performance, while NbOsBi \cite{Boutaiba_2026}, HfIrBi \cite{Hu_2023}, and LuAuPb \cite{Singh_2016} have also shown promising thermoelectric properties. These subsequent independent studies provide retrospective support for the ability of the ML models to identify chemically meaningful thermoelectric candidates beyond the original training space.

The Mg-based compositions predicted by the XGB and NN models, namely MgAgBi, MgAuBi, and MgAuSb, were all found to be metallic in our DFT calculations and were therefore not investigated further as conventional semiconducting half-Heusler thermoelectrics. Consequently, although the XGB and RF models successfully identified NbOsBi and HfIrBi, their predictions were less consistent overall. By contrast, compositions identified simultaneously by multiple ML models, particularly those predicted by SISSO and NN, show stronger correspondence with the high-performing thermoelectric half-Heusler compounds subsequently reported in the literature. Overall, the retrospective comparison indicates that all four ML architectures are capable of identifying chemically meaningful candidates beyond the training data, although their extrapolative consistency differs across architectures.

CeCoBi and CeIrBi were selected for direct first-principles validation because, among the candidates identified in Table~5, they were the only compositions that had neither been experimentally reported nor previously investigated computationally at the time of our screening. Importantly, Ce is absent from the training dataset: no Ce-containing compound was used for feature selection, hyperparameter optimization or model training. Because the models describe elements through their physical descriptors rather than their identity, the high $zT$ predicted for CeCoBi and CeIrBi rests entirely on the transfer of the learned descriptor--$zT$ relationships to an element not encountered during training, making these compounds a demanding test of extrapolation to new chemistry. For CeIrBi, the test is further strengthened by the fact that Ir appears in the training data only as a minor B-site dopant (NbFe$_{1-x}$Ir$_{x}$Sb, $x \leq 0.08$), never as the host B-site element. The DFT results in the following section, which presents the electronic structures and thermoelectric properties of CeCoBi and CeIrBi together with a comparison with their ML-predicted $zT$ values, therefore demonstrate the extrapolative capacity of the models for this case.We note, however, that in these compounds Ce is the only element entirely absent from the training data; we do not claim out-of-distribution generalizability to compositions in which several or all constituent
elements are unseen. The following section presents the DFT-calculated electronic structures and thermoelectric properties of CeCoBi and CeIrBi, including comparison with their ML-predicted $zT$ values.

\begin{table}[htbp]
\centering
\caption{Top-ranked half-Heusler compositions identified from the Legrain \textit{et al.} (2018)\cite{Legrain_2018}  candidate set using the present ML models. The stability values correspond to the predicted probability of adopting the half-Heusler structure, and $\zTpr$ denotes our predicted thermoelectric figure of merit at 673~K.}
\label{tab:HTS_Legrain}
\begin{tabular}{l l c c}
\hline
Model & Composition & Stability & $\zTpr$ (673 K) \\
\hline
SISSO & \textbf{CeCoBi} & 0.611 & 0.370 \\
      & NbFeBi          & 0.699 & 0.369 \\
      & \textbf{CeIrBi} & 0.663 & 0.331 \\
      & LuPdBi          & 0.831 & 0.300 \\
\hline
NN    & TaFeBi & 0.658 & 0.422 \\
      & HfCoBi & 0.844 & 0.405 \\
      & NbFeBi & 0.699 & 0.397 \\
      & LuAuPb & 0.708 & 0.370 \\
\hline
XGB   & MgAgBi & 0.612 & 0.451 \\
      & MgAuBi & 0.609 & 0.422 \\
      & MgAuSb & 0.775 & 0.381 \\
      & HfIrBi & 0.776 & 0.353 \\
\hline
RF    & MgAuSb & 0.775 & 0.351 \\
      & MgAuBi & 0.609 & 0.343 \\
      & MgAgBi & 0.612 & 0.339 \\
      & NbOsBi & 0.628 & 0.304 \\
\hline
SISSO+NN+XGB+RF
      & NbFeBi & 0.699 & 0.330 \\
      & HfCoBi & 0.844 & 0.329 \\
      & HfIrBi & 0.776 & 0.328 \\
      & TaFeBi & 0.658 & 0.327 \\
\hline
\end{tabular}
\end{table}

\subsubsection{3.4.2 DFT validation of CeCoBi and CeIrBi}

\subsubsection{i. Structural relaxation of CeCoBi and CeIrBi}

The crystal structures of CeCoBi and CeIrBi were fully relaxed within density functional theory (as detailed in the Methods section), and the optimized structural parameters are summarized in Table~6. Both compounds crystallize in the non-centrosymmetric cubic half-Heusler (MgAgAs-type) structure with space group $F\bar{4}3m$ (No.~216). The optimized lattice constants, bond lengths, and unit-cell volumes are systematically larger than those of the corresponding ZrCoBi \cite{MP_2020} and HfIrBi\cite{Hu_2023} reference compounds owing to the larger atomic radius of Ce relative to Zr and Hf. Furthermore, the structural parameters of CeIrBi exceed those of CeCoBi because of the larger atomic size of Ir compared with Co.

\begin{table}[htbp]
\centering
\caption{Calculated structural parameters of $AB$Bi ($A$: Ce, Hf, Zr; $B$: Co, Ir) half-Heusler alloys.}
\label{tab:structural_parameters_CeCoBi_CeIrBi}
\begin{tabular}{lccccc}
\hline
Material & $a$ (\AA) & $d_{\mathrm{A-B}}$ (\AA) & $d_{\mathrm{A-Bi}}$ (\AA) & $V$ (\AA$^3$) & $E_g$ (eV) \\
\hline
CeCoBi [This work] & 6.527 & 2.826 & 3.264 & 278.06 & 0.2 \\
ZrCoBi \cite{MP_2020} & 6.190 & 2.700 & 3.120 & 236.96 & 0.98 \\
CeIrBi [This work] & 6.763 & 2.929 & 3.381 & 309.38 & 0 \\
HfIrBi \cite{Hu_2023} & 6.460 & 2.798 & 3.230 & 269.58 & 0 \\
\hline
\end{tabular}
\end{table}

Although the two compounds adopt the same crystal structure, their electronic ground states differ significantly. CeCoBi is predicted to be a narrow-gap semiconductor with a calculated band gap of $\sim$0.2~eV, which is considerably smaller than that of ZrCoBi ($\sim$0.98~eV) and among the smallest reported for Co-based half-Heusler compounds~\cite{ Yang_2022, surucu_2020, matth_2023}. Such a narrow band gap is generally advantageous for thermoelectric applications because it facilitates higher carrier concentrations while maintaining a sufficiently large Seebeck coefficient~\cite{gibbs_2015}. In contrast, CeIrBi exhibits a zero band gap, consistent with previous calculations for HfIrBi~\cite{Hu_2023}, indicating a semimetallic electronic structure. Although semimetals are traditionally considered less attractive for thermoelectric applications because of their high intrinsic carrier concentrations, recent theoretical and experimental studies have demonstrated that semimetals possessing strong electron--hole asymmetry and favorable band convergence can simultaneously achieve large Seebeck coefficients and high power factors~\cite{markov_2019}. The detailed electronic structures of both compounds are therefore investigated in the following sections.

It should be emphasized that the primary objective of the present first-principles calculations is not to provide a complete \textit{ab initio} characterization of CeCoBi and CeIrBi, but rather to validate the thermoelectric potential predicted by the SISSO models through our investigation of their electronic structures and transport properties. A comprehensive first-principles assessment of the materials' stability---including phase stability (convex-hull analysis), mechanical stability (elastic constants), and dynamical stability (phonon dispersion)---together with a rigorous determination of the thermoelectric figure of merit through anharmonic lattice thermal conductivity calculations, will be presented in a subsequent study.

\subsubsection{ii. Electronic structure of CeCoBi and CeIrBi}

The total and projected density of states (DOS/PDOS), together with the electronic band structures, were calculated to elucidate the electronic states governing the transport properties of CeCoBi and CeIrBi. The DOS/PDOS and band structures are presented in Figures~9a, 9b and Figures~9c, 9d, respectively. For CeCoBi, the valence band is dominated by Co-$3d$ states with secondary contributions from Bi-$6p$ orbitals and only a minor contribution from Ce-$5d$ states, giving rise to three prominent DOS peaks located at approximately $-2.6$, $-1.1$, and $-0.2$~eV below the Fermi level. This orbital character is consistent with that reported for Co-based 18-valence-electron half-Heusler compounds \cite{Pati_2024} such as TiCoBi, ZrCoBi, and HfCoBi, where the transition-metal $d$ states dominate the occupied electronic states. Above the Fermi level, the conduction band also remains predominantly Co-$3d$ in character, while the Bi-$6p$ and Ce-$5d$ orbitals contribute with comparable intensities, both remaining substantially smaller than the Co-$3d$ contribution. Unlike conventional 18-VEC half-Heusler compounds, where the conduction-band edge is typically dominated by the A-site $d$ orbitals \cite{Pati_2024}, CeCoBi exhibits only a modest participation of Ce-$5d$ states. This behavior suggests relatively weak hybridization between the Ce-$5d$ orbitals and the anionic [CoBi]$^{4-}$ framework, resulting in the low-energy electronic structure being governed primarily by Co-$3d$ states hybridized with Bi-$6p$ orbitals. The Ce-$4f$ states (not shown) are located well above the Fermi level and therefore do not contribute to the occupied valence band. The calculated band structure confirms that CeCoBi is an indirect-bandgap semiconductor, with the valence-band maximum located at L and the conduction-band minimum at $\Gamma$. Two nearly degenerate valence bands occur at L (multiplicity = 4), while an additional valence-band maximum is present at X (multiplicity = 3) within a small energy difference ($<0.1$~eV) shown by the red line. The resulting valley degeneracy is $N_v = (2 \times 4) + (1 \times 3) = 11$. The small energy separation between the L- and X-derived valence-band maxima enables both valleys to participate in hole transport, increasing the density-of-states effective mass and favoring a large Seebeck coefficient.

In contrast, CeIrBi exhibits a pseudo-gap with a finite density of states at the Fermi level, consistent with its zero-gap semi-metallic electronic structure. The upper valence band is dominated by Ir-$5d$ states with significant Bi-$6p$ contributions, whereas the Ce-$5d$ contribution remains comparatively small. Above the Fermi level, the conduction band continues to be dominated by Ir-$5d$ states, while Ce-$5d$ and Bi-$6p$ orbitals contribute with comparable intensities, both remaining substantially smaller than the Ir-$5d$ contribution. The Ir-$6s$ and Bi-$6s$ states make negligible contributions near the Fermi level. The weak participation of Ce-$5d$ states indicates relatively weak hybridization between the Ce-$5d$ orbitals and the anionic [IrBi]$^{4-}$ framework. The electronic band structure confirms that CeIrBi is a zero-gap semi-metal in which the valence and conduction bands touch at the $\Gamma$ point with predominantly quadratic dispersion. Within 0.1~eV below the Fermi level (red line), two nearly degenerate valence bands occur at both $\Gamma$ (multiplicity = 1) and L (multiplicity = 4), giving a total valley degeneracy of $N_v = (2 \times 1) + (2 \times 4) = 10$.

\begin{figure}[htbp]
    \centering
    \includegraphics[width=\textwidth]{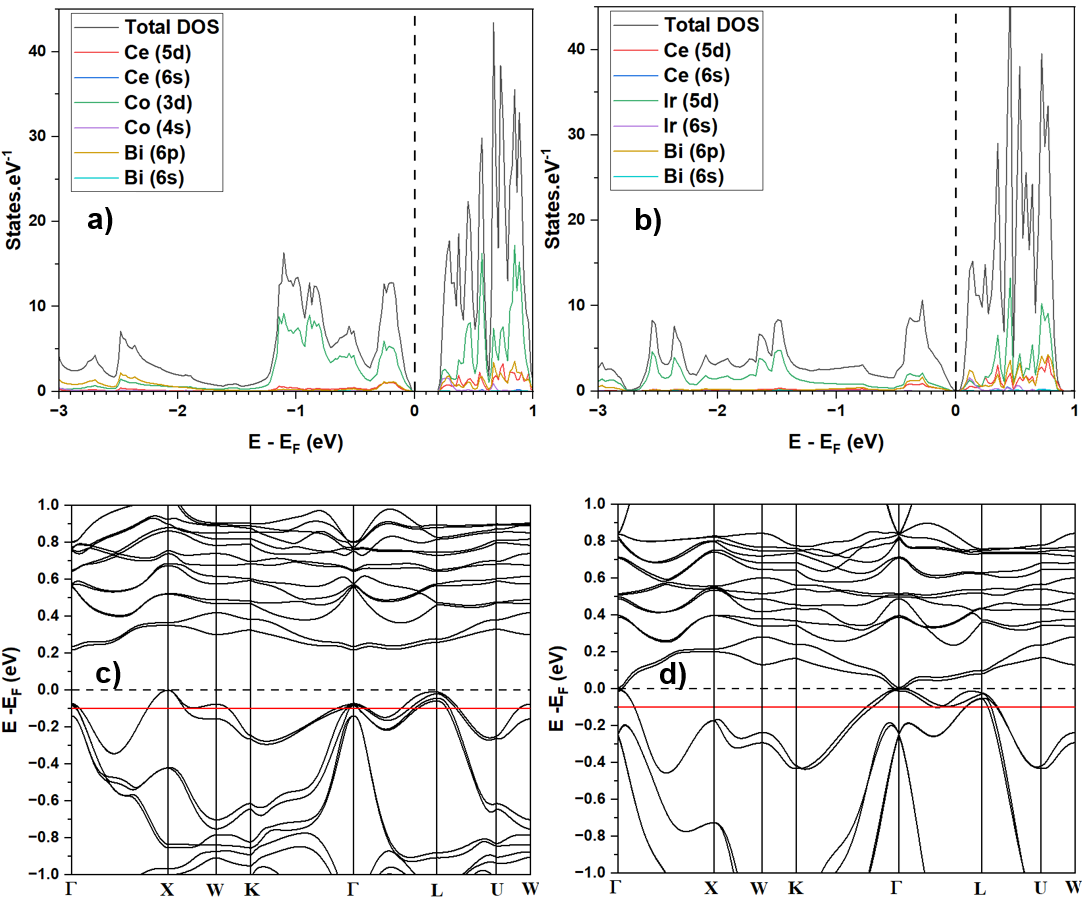}
    \caption{Calculated total and projected density of states of (a) CeCoBi and (b) CeIrBi; calculated electronic band structures of (c) CeCoBi and (d) CeIrBi.}
    \label{fig:electronic_structure}
\end{figure}

Overall, CeCoBi and CeIrBi have different electronic mechanisms for achieving favorable thermoelectric properties. CeCoBi benefits from an indirect-gap semiconducting electronic structure with pronounced valence-band convergence and $N_v = 11$, whereas CeIrBi exhibits a pseudo-gapped semi-metallic electronic structure with $N_v = 10$. Despite these differences, both compounds share a common orbital characteristic: the conduction-band edge is dominated by the transition-metal $d$ states (Co-$3d$ or Ir-$5d$), while the Bi-$6p$ and Ce-$5d$ orbitals contribute with comparable but considerably smaller intensities. This contrasts with conventional Ti-, Zr-, and Hf-based 18-valence-electron half-Heusler compounds, where the A-site $d$ orbitals dominate the conduction band \cite{Pati_2024}. The consistently weak participation of Ce-$5d$ states in both compounds indicates relatively weak hybridization between the Ce-$5d$ orbitals and the anionic [TBi]$^{4-}$ (T = Co, Ir) framework, suggesting that this is an intrinsic electronic characteristic of these Ce-based half-Heusler compounds.

\subsubsection{iii. Thermoelectric Properties of CeCoBi and CeIrBi}

The temperature dependence of the Seebeck coefficient ($S$), electrical conductivity ($\sigma$), and power factor (PF) of intrinsic CeCoBi and CeIrBi, calculated within the constant relaxation time approximation (CRTA) using a relaxation time of $\tau = 10^{-14}$~s, is shown in Figure~6. Within the CRTA, $S$ is independent of the relaxation time, whereas $\sigma$ and PF scale linearly with $\tau$. The two compounds exhibit distinct transport behavior that reflects their contrasting electronic structures. CeCoBi, an indirect-gap semiconductor, displays a characteristic monotonic decrease in the Seebeck coefficient from approximately $390~\mu$V~K$^{-1}$ at 300~K to $216~\mu$V~K$^{-1}$ at 973~K, exceeding the values reported for several pristine Co-based half-Heusler compounds \cite{Pati_2024, Yang_2022}. This behavior originates from the pronounced valence-band convergence and the large valley degeneracy ($N_v = 11$). In contrast, the semimetallic CeIrBi exhibits a nearly temperature-independent Seebeck coefficient, increasing slightly from $167~\mu$V~K$^{-1}$ at 300~K to a maximum of $179~\mu$V~K$^{-1}$ near 473~K before gradually decreasing to $160~\mu$V~K$^{-1}$ at 973~K. Despite the absence of a true band gap, these values are comparable to those reported for high-performance semimetallic thermoelectric materials such as HgTe \cite{markov_2018}and exceed previous theoretical calculations for HfIrBi \cite{Hu_2023}.

The electrical conductivity increases monotonically with temperature for both compounds because of the increasing thermally excited carrier concentration. However, the semi-metallic electronic structure of CeIrBi results in substantially higher conductivity over the entire temperature range. Specifically, the calculated conductivity of CeCoBi increases from approximately $3.35 \times 10^{3}$~S~m$^{-1}$ at 300~K to $6.37 \times 10^{4}$~S~m$^{-1}$ at 673~K, reaching $1.75 \times 10^{5}$~S~m$^{-1}$ at 973~K. In comparison, CeIrBi exhibits significantly larger values of $6.40 \times 10^{4}$, $2.63 \times 10^{5}$, and $4.40 \times 10^{5}$~S~m$^{-1}$ at 300, 673, and 973~K, respectively. Thus, although CeCoBi possesses the larger Seebeck coefficient, the substantially higher electrical conductivity of CeIrBi leads to superior electronic transport performance. The combined behavior of $S$ and $\sigma$ results in continuously increasing power factors for both materials. At the technologically relevant prediction temperature of 673~K, CeCoBi and CeIrBi exhibit intrinsic power factors of 5.26 and 8.39~mW~m$^{-1}$~K$^{-2}$, respectively, increasing further to 8.16 and 11.4~mW~m$^{-1}$~K$^{-2}$ at 973~K. The calculated PF of CeCoBi is comparable to that of optimized half-Heusler thermoelectric materials at elevated temperatures \cite{zhu_2019, fu_2015}, whereas the PF of CeIrBi is comparable to those reported for HfIrBi and the benchmark semimetal HgTe \cite{markov_2018}. These results demonstrate that both compounds possess excellent intrinsic electronic transport characteristics despite their fundamentally different electronic structures.

\begin{figure}[htbp]
    \centering
    \includegraphics[width=0.95\textwidth]{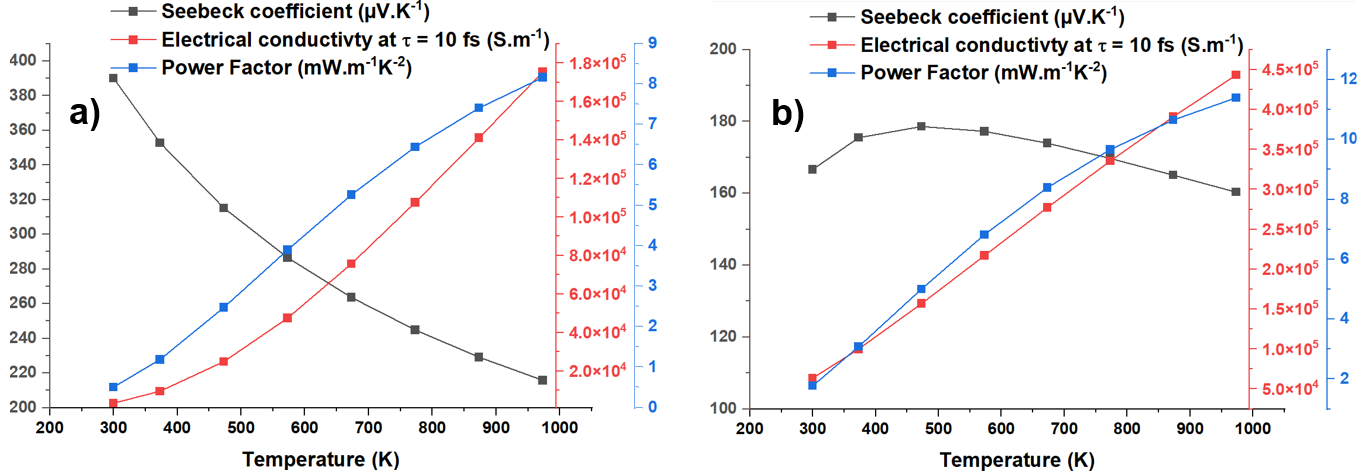}
    \caption{Calculated Seebeck coefficient, electrical conductivity ($\tau = 10^{-14}$~s), and power factor as a function of temperature for CeCoBi and CeIrBi.}
    \label{fig:TE_properties}
\end{figure}

The present first-principles calculations therefore provide independent support for the thermoelectric potential predicted by the SISSO models. Although a quantitative determination of the thermoelectric figure of merit $zT$ requires the lattice thermal conductivity, which is obtained from computationally demanding anharmonic finite-displacement calculations, the calculated electronic structures and transport properties provide important original evidence for the electronic contribution to thermoelectric performance. In particular, CeCoBi combines a high Seebeck coefficient with pronounced valence-band convergence ($N_v = 11$), whereas CeIrBi achieves a favorable balance between Seebeck coefficient and electrical conductivity through its pseudo-gapped semi-metallic electronic structure ($N_v = 10$). The resulting intrinsic power factors of 5.26~mW~m$^{-1}$~K$^{-2}$ for CeCoBi and 8.39~mW~m$^{-1}$~K$^{-2}$ for CeIrBi at 673~K are consistent with their identification by the SISSO machine-learning models as two promising half-Heusler thermoelectric compounds, with predicted $zT$ values of 0.37 and 0.33, respectively. While these DFT calculations do not constitute a complete first-principles validation of the predicted $zT$ values because the lattice thermal conductivity has not yet been explicitly calculated, they provide additional physical support for the SISSO predictions by demonstrating that both compounds possess favorable electronic structures and strong intrinsic electronic transport characteristics. A complete first-principles assessment of their thermoelectric performance, including anharmonic lattice thermal conductivity and the resulting quantitative $zT$, will be pursued in future work.

\subsubsection{3.4.3 High-throughput screening (HTS) of doped compositional space}

Following the evaluation of the ML models on previously unexplored undoped 
half-Heusler compositions, including retrospective comparison with subsequent 
literature reports and our first-principles assessment of CeCoBi and 
CeIrBi, we applied the framework to the substantially larger space of doped 
half-Heusler compositions. As described above, the doped chemical space is 
defined by 631 stability-filtered parent $ABC$ hH matrices combined with 25, 
14, and 9 candidate dopants at the $A$, $B$, and $C$ sites, respectively, 
yielding approximately $2.46\times10^{6}$ distinct elemental substitution 
combinations. Each combination was further expanded over the 
substitution-concentration space, varying the dopant concentration in two 
regimes: a fine grid with a step of 0.01 from 0 to 0.1 and a coarser grid with 
a step of 0.05 from 0.1 to 0.5 at each doped site, and screened using the 
Slater--Pauling VEC criterion described above. The highest-ranked candidates 
obtained from within-technique and across-technique inverse-error-weighted 
ensemble predictions are listed in Table~7. These candidates should be regarded 
as computationally prioritized compositions rather than experimentally 
established high-$zT$ materials.


The use of ensemble predictions is motivated by the complementary behavior of the different ML architectures. A collection of models trained on different material-level subsets can provide more stable predictions than an individual model by reducing sensitivity to a particular training realization \cite{naftaly_1997}. In the present framework, ensemble averaging also captures complementary local trends learned by SISSO, NN, XGB, and RF. Furthermore, because the screening objective is focused on a technologically relevant temperature of $\ThtsK$, the predictions are evaluated at this fixed temperature using the two-stage inverse-error-weighting procedure described above. This weighting gives greater importance to folds and techniques that demonstrate better predictive performance at the target temperature while retaining contributions from the different ML architectures. The spread among the predictions of SISSO, NN, XGB, and RF provides an additional qualitative indication of model disagreement and, consequently, of epistemic uncertainty. Candidates for which the different models yield similar predictions can therefore be regarded as more robust than candidates associated with large inter-model deviations. This spread is used only as a qualitative indicator and is not interpreted as a formally calibrated predictive uncertainty interval.

Direct first-principles validation of the entire doped compositional space is 
computationally prohibitive. In particular, substitutional disorder and the 
associated local structural relaxation generally require explicit supercell 
calculations, making brute-force DFT screening impractical for a chemical space 
containing approximately $2.46\times10^{6}$ elemental substitution combinations, 
each further expanded over a fine and coarse dopant-concentration grid. The 
purpose of the present ML-based HTS is therefore to reduce this large chemical 
space to a tractable set of candidates for subsequent theoretical and 
experimental investigation. The resulting shortlists provide a model-ensemble-based 
prioritization rather than a guarantee of high thermoelectric performance.

Database-informed stability filtering further reduces the search space by imposing the thermodynamic and electronic criteria described above. In particular, $ABC$ matrices for which hull-distance information was unavailable in either OQMD or the Materials Project were deliberately excluded. Such systems may contain potentially interesting compositions, but their inclusion would require additional first-principles stability calculations and would substantially increase the computational cost of the screening. Similarly, the database-specific hull-distance cutoffs used here should not be interpreted as absolute criteria for experimental synthesizability. Materials located above the convex hull may nevertheless be accessible through non-equilibrium synthesis routes or kinetic stabilization. The stability filters are therefore intended to provide a practical and reproducible reduction of the chemical space rather than a definitive prediction of experimental phase formation.

Several of the highest-ranked predictions involve established half-Heusler matrices with a demonstrated thermoelectric record. For example, NbFeSb \cite{mohamed_2020}, TaFeSb \cite{zhu_2019}, and ZrCoBi \cite{zhu_2018} have previously been investigated with different dopants and compositions and have demonstrated favorable thermoelectric performance. Their recurrence among the high-ranked candidates provides a materials-level consistency check of the screening framework. At the same time, the screening identifies less-explored matrices, including HfNiSb, TiNiGe, and ZrNiBi, which provide potential directions for further investigation. These predictions should, however, be regarded as computational hypotheses requiring independent stability, electronic-structure, and experimental assessment.

A pronounced chemical trend in the SISSO and NN predictions is the frequent occurrence of light A-site dopants, particularly Li and Mg, with Ca also appearing in some high-ranked compositions. The substantial size and mass mismatch between these dopants and heavier A-site host atoms provides a physically motivated route for enhanced point-defect scattering and, potentially, reduced lattice thermal conductivity ($\kappa_L$). This interpretation is consistent with the importance of A-site descriptors identified in the SISSO and SHAP analyses. However, phonon-scattering rates have not been explicitly calculated for the individual HTS candidates, and therefore this mechanism should be regarded as a physically motivated hypothesis rather than a quantitatively demonstrated effect. Moreover, light-element substitution may introduce additional complications, including limited solid solubility, interstitial defect formation, and secondary-phase formation. These effects cannot be resolved from the present ML screening alone and require subsequent first-principles and experimental investigation.

The ML architectures also exhibit partially complementary chemical trends. SISSO and NN preferentially identify several Li- and Mg-containing compositions, whereas XGB and RF yield different substitutional motifs and, in some cases, place greater emphasis on other regions of the doped chemical space. This complementarity provides a materials-level rationale for using across-technique ensemble averaging rather than relying on a single ML architecture. At the same time, the differences between the individual model predictions emphasize the importance of treating the ensemble as a prioritization framework rather than assuming that every high-ranked candidate has the same level of predictive confidence.

The experimental validation of these predictions is being pursued within an ongoing DFG--ANR CombiHeusler project using thin-film material libraries (TFML) \cite{talley_2019}. Co-sputtering of the constituent elements can enable the synthesis and screening of compositionally diverse thin-film libraries, allowing promising candidates to be characterized before targeted bulk synthesis is attempted. Such an approach can substantially reduce the experimental effort required to investigate the large number of compositions generated by the HTS. Importantly, experimental measurements would also provide new training data for subsequent refinement of the ML models.

In this context, an active-learning strategy provides a natural extension of the present workflow. Compositions for which experimental measurements exhibit large discrepancies from the ML predictions can be incorporated into the training dataset, followed by retraining and reevaluation of the models. Iterative selection of informative compositions can progressively enrich the dataset in regions of chemical space where the present models are most uncertain or inaccurate. Such active-learning approaches are particularly attractive for materials science because experimental data generation is both costly and time-consuming \cite{kusne_2020}. By prioritizing experimentally informative compositions rather than sampling the chemical space uniformly, the combination of TFML-based experimental validation and active learning could reduce the experimental workload while improving the generalizability of the models. This integrated ML--HTS--experimental workflow will be explored in future work.

\begin{table}[htbp]
\centering
\small
\caption{Highest-ranked HTS candidates at 673~K obtained from within-technique and
across-technique inverse-error-weighted ensemble predictions.}
\label{tab:hts}
\begin{tabular}{lll}
\hline
Technique & Composition & $zT_{\mathrm{pr}}$ \\
\hline
SISSO$^{a}$ & $\mathrm{Hf}_{0.55}\mathrm{Li}_{0.45}\mathrm{Ni}_{1-x}\mathrm{Zn}_{x}\mathrm{Sb}_{1-y}\mathrm{Bi}_{y}$ ($0.01 \le x, y \le 0.5$) & 1.24 \\
            & $\mathrm{Ta}_{0.55}\mathrm{Li}_{0.45}\mathrm{Co}_{1-x}\mathrm{Fe}_{x}\mathrm{Sb}_{1-y}\mathrm{Sn}_{y}$ ($x, y \ge 0.01$, $x + y < 0.2$) & 1.19 \\
            & $\mathrm{Nb}_{0.5}\mathrm{Li}_{0.5}\mathrm{Co}_{1-x}\mathrm{Pt}_{x}\mathrm{Sb}_{1-y}\mathrm{Sn}_{y}$ ($0.01 \le y < x \le 0.5$) & 1.18 \\
            & $\mathrm{Ti}_{0.5}\mathrm{Mg}_{0.5}\mathrm{Ni}_{1-x}\mathrm{Zn}_{x}\mathrm{Ge}_{1-y}\mathrm{Si}_{y}$ ($0.01 \le x, y \le 0.5$) & 1.08 \\
\hline
NN  & $\mathrm{Ta}_{0.75}\mathrm{Li}_{0.25}\mathrm{Fe}_{0.95}\mathrm{Ag}_{0.05}\mathrm{Sb}$ & 0.77 \\
    & $\mathrm{Nb}_{0.8}\mathrm{Li}_{0.2}\mathrm{Fe}_{0.95}\mathrm{Cu}_{0.05}\mathrm{Sb}$ & 0.75 \\
    & $\mathrm{Zr}_{0.75}\mathrm{Li}_{0.25}\mathrm{Co}_{0.75}\mathrm{Cd}_{0.25}\mathrm{Bi}$ & 0.74 \\
    & $\mathrm{Ta}_{0.75}\mathrm{Mg}_{0.25}\mathrm{Co}_{0.95}\mathrm{Cu}_{0.05}\mathrm{Sb}_{0.95}\mathrm{Ge}_{0.05}$ & 0.74 \\
\hline
XGB & $\mathrm{Ta}_{0.85}\mathrm{Ti}_{0.15}\mathrm{Fe}\mathrm{Sb}_{0.95}\mathrm{Al}_{0.05}$ & 0.68 \\
    & $\mathrm{Nb}_{0.85}\mathrm{Ca}_{0.15}\mathrm{Fe}\mathrm{Sb}$ & 0.64 \\
    & $\mathrm{Zr}\mathrm{Co}_{0.95}\mathrm{Pt}_{0.05}\mathrm{Bi}_{0.55}\mathrm{Sn}_{0.45}$ & 0.64 \\
    & $\mathrm{Zr}_{0.95}\mathrm{Hf}_{0.05}\mathrm{Ni}\mathrm{Bi}_{0.55}\mathrm{Sn}_{0.45}$ & 0.63 \\
\hline
RF  & $\mathrm{Ta}_{0.85}\mathrm{Ti}_{0.15}\mathrm{Fe}\mathrm{Sb}_{0.95}\mathrm{Bi}_{0.05}$ & 0.74 \\
    & $\mathrm{Zr}\mathrm{Co}_{0.95}\mathrm{Cu}_{0.05}\mathrm{Bi}_{0.5}\mathrm{Sn}_{0.5}$ & 0.69 \\
    & $\mathrm{Nb}_{0.85}\mathrm{Ti}_{0.15}\mathrm{Fe}\mathrm{Sb}_{0.95}\mathrm{Bi}_{0.05}$ & 0.65 \\
    & $\mathrm{Zr}\mathrm{Ni}_{0.95}\mathrm{Cu}_{0.05}\mathrm{Bi}_{0.5}\mathrm{Sn}_{0.5}$ & 0.61 \\
\hline
Ensemble$^{b}$ & $\mathrm{Ta}_{0.85}\mathrm{Ca}_{0.15}\mathrm{Fe}\mathrm{Sb}$ & 0.68 \\
                & $\mathrm{Zr}_{0.7}\mathrm{Li}_{0.3}\mathrm{Co}_{0.95}\mathrm{Zn}_{0.05}\mathrm{Bi}_{0.8}\mathrm{Sn}_{0.2}$ & 0.65 \\
                & $\mathrm{Nb}_{0.85}\mathrm{Ca}_{0.15}\mathrm{Fe}\mathrm{Sb}$ & 0.63 \\
                & $\mathrm{Zr}\mathrm{Ni}_{0.95}\mathrm{Cu}_{0.05}\mathrm{Bi}_{0.5}\mathrm{Sn}_{0.5}$ & 0.60 \\
\hline
\end{tabular}

\smallskip
\begin{minipage}{0.95\linewidth}
\footnotesize
$^{a}$The final SISSO descriptors do not include $x_{B'}$ and $x_{C'}$, so the predicted $zT$
is independent of these concentrations. The ranges given are those on the screening grid for
which the Slater--Pauling criterion ($17 < \mathrm{VEC} < 19$) is satisfied with nonzero dopant
concentrations.
$^{b}$Across-technique ensemble of SISSO, NN, XGB and RF.
\end{minipage}
\end{table}
\section{4. Conclusion}

In this work, we developed a robust and generalizable machine-learning workflow for predicting and screening ABC half-Heusler (hH) thermoelectric materials. A PCA-based material-wise train/test splitting strategy, combined with a correlation-based feature-filtering procedure integrating Bayesian hyperparameter optimization with $k$-best selection, was used across Random Forest, XGBoost, and Neural Network models, reducing the test RMSE by 16–27\% relative to the full feature set. This was complemented by the interpretable SISSO symbolic-regression method. SHAP and SISSO analyses identified temperature as the dominant driver of $zT$, with the A-site dopant concentration and heat of vaporization emerging as the most important material-dependent descriptors, underscoring the central role of the A-site in hH thermoelectric performance. Extrapolative capacity was assessed through three complementary approaches: evaluation on previously unexplored undoped compositions, identified by Legrain \textit{et al.} (2018) \cite{Legrain_2018}, several of which were subsequently supported by independent literature; recovery of literature-reported optimal $zT$ values for ten established hH systems, with SISSO showing the closest agreement; and new DFT calculations for two previously unreported compositions, CeCoBi and CeIrBi, which yielded favorable intrinsic power factors of 5.26 and 8.39~mW~m$^{-1}$~K$^{-2}$ at 673~K, respectively, supporting their promise as good hH thermoelectric candidates. As a practical demonstration of the framework's utility, we applied it to the substantially larger doped hH space (approximately $2.46\times10^{6}$ elemental substitution combinations across 631 thermodynamically screened parent matrices), identifying several promising, previously unstudied compositions for future first-principles and experimental investigation. Overall, this work demonstrates that combining rigorous data partitioning, principled feature filtering, complementary ML architectures, and targeted extrapolative validation -- rather than reliance on conventional performance metrics alone -- provides a practical and generalizable framework for navigating the large chemical space of hH thermoelectrics. Coupling this framework with thin-film material libraries and active-learning strategies offers a scalable path toward accelerating the discovery of high-performance thermoelectric materials.

\section{Data and Software Availability}

The PCA-based train/test splitting code and the fold-wise assignment of material IDs used in this work are publicly available at
\url{https://src.koda.cnrs.fr/icgm-thermoelectrics-ml}.
The complete processed dataset and trained model files will be made available in the same repository once the dataset has been finalized and the experimental validation of the screened candidates has been completed, respectively. The repository is maintained and will be updated as additional reproducibility materials become available.

\bibliography{References}

\section{Acknowledgement}

We acknowledge the financial support from the French Agence Nationale de la Recherche (ANR) under the ANR-DFG project CombiHeusler (ANR-24-CE92-0053-01).

\section{Author Contribution declaration} 

S.A.: Writing—Original draft preparation, Visualization, Methodology, Investigation, Data curation, Conceptualization; A.M.: Writing—Original draft preparation, Visualization, Methodology, Investigation, Data curation; P.J.: Methodology,  Conceptualization, Supervision, Writing-Reviewing and Editing, Validation

\section{Competing Interest declaration} 

The authors declare no competing interests.

\end{document}